\documentclass[12pt]{article}
\usepackage{amsmath,bm}
\usepackage{mathrsfs}
\usepackage{amssymb}
\makeatletter \@addtoreset{equation}{section} \makeatother
\renewcommand{\theequation}{\thesection.\arabic{equation}}
\newcommand{\ba}{\begin{array}}
\newcommand{\ea}{\end{array}}
\newcommand{\beq}{\begin{equation}}
\newcommand{\eeq}{\end{equation}}
\newcommand{\bea}{\begin{eqnarray}}
\newcommand{\eea}{\end{eqnarray}}

\def\bce{\begin{center}}
\def\ece{\end{center}}

\def\nonu{\nonumber}

\def\pa{\partial}

\def\be{\beta}

\def\de{\delta}

\def\si{\sigma}

\def\eps6{{\displaystyle \mathop{\epsilon}^{6}}{}}
\def\g6{{\displaystyle \mathop{g}^{6}}{}}

\def\nab6{{\displaystyle \mathop{\nabla}^{6}}{}}

\newcommand\T{\rule{0pt}{3ex}} 
\newcommand\B{\rule[-1.6ex]{0pt}{0pt}}

\def\0{{\sst{(0)}}}
\def\1{{\sst{(1)}}}
\def\2{{\sst{(2)}}}
\def\3{{\sst{(3)}}}
\def\4{{\sst{(4)}}}
\def\5{{\sst{(5)}}}
\def\6{{\sst{(6)}}}
\def\7{{\sst{(7)}}}
\def\8{{\sst{(8)}}}

\def\ba{\begin{array}}
\def\ea{\end{array}}
\def\beq{\begin{equation}}
\def\eeq{\end{equation}}
\def\be{\begin{equation}}
\def\ee{\end{equation}}
\def\braket#1{\left\langle #1 \right\rangle}

\def\eps{\epsilon}

\def\ba{\begin{array}}
\def\ea{\end{array}}
\def\beq{\begin{equation}}
\def\eeq{\end{equation}}
\def\be{\begin{equation}}
\def\ee{\end{equation}}
\def\braket#1{\left\langle #1 \right\rangle}

\def\eps{\epsilon}

\def\eps6{{\displaystyle \mathop{\epsilon}^{6}}{}}

\def\nab6{{\displaystyle \mathop{\nabla}^{6}}{}}

\newcommand{\bean}{\begin{eqnarray*}}
\newcommand{\eean}{\end{eqnarray*}}

\begin{document}

\baselineskip=18pt
\renewcommand{\theequation}
{\arabic{section}\mbox{.}\arabic{equation}}


\thispagestyle{empty} \addtocounter{page}{-1}
   \begin{flushright}
\end{flushright}

\vspace*{1.3cm}
  
\centerline{ \Large \bf
The Small ${\cal N}=4$  Superconformal ${\cal W}_{\infty}$ Algebra
}
\vspace*{1.5cm}
\centerline{{\bf  Changhyun Ahn$^\dagger$},
  {\bf Matthias R. Gaberdiel$^\star$} and {\bf Man Hea Kim$^{\dagger\star\ast} \footnote
    {Address after April 1, 2020:
 Asia Pacific Center for Theoretical Physics, Pohang 37673, Korea}$}
} 
\vspace*{1.0cm} 
\centerline{\it 
$\dagger$ Department of Physics, Kyungpook National University, Taegu
41566, Korea} 
\vspace*{0.5cm}
\centerline{\it 
  $\star$ Institut f$\ddot{u}$r Theoretische Physik,
  ETH Zurich, 8093 Z$\ddot{u}$rich, Switzerland}
\vspace*{0.5cm}
\centerline{\it 
  $\ast$
  Asia Pacific Center for Theoretical Physics, Pohang 37673, Korea }
\vspace*{0.8cm} 
\centerline{\tt ahn@knu.ac.kr,  gaberdiel@itp.phys.ethz.ch, manhea.kim@apctp.org 
} 
\vskip2cm

\centerline{\bf Abstract}
\vspace*{0.5cm}

The symmetric orbifold of $\mathbb{T}^4$ is the CFT dual of string theory on AdS$_3\times {\rm S}^3 \times \mathbb{T}^4$ with minimal NS-NS flux. We study its symmetry algebra and provide evidence that  it does not have any deformation parameter. This suggests that the symmetric orbifold is (at least locally) the most symmetrical CFT in its moduli space.

\baselineskip=18pt
\newpage
\renewcommand{\theequation}
{\arabic{section}\mbox{.}\arabic{equation}}

\tableofcontents

\section{Introduction}

It was recently realised that string theory on ${\rm AdS}_3 \times {\rm S}^3 \times \mathbb{T}^4$ with one unit of NS-NS flux is exactly dual to the symmetric orbifold of $\mathbb{T}^4$ \cite{Gaberdiel:2018rqv,Eberhardt:2018ouy,Eberhardt:2019ywk}. At this point in moduli space the string theory is `tensionless' since the size of the AdS background (which is proportional to the flux) is minimal. In turn, the dual CFT is therefore expected to describe the analogue of free super Yang-Mills theory \cite{Sundborg:2000wp,Witten,Mikhailov:2002bp}, and one should expect a subsector of the theory to exhibit a higher spin/CFT duality as in \cite{Klebanov:2002ja}, see also \cite{Sezgin:2003pt,Giombi:2009wh,Giombi:2010vg} for subsequent work. 

This expectation is indeed borne out \cite{Gaberdiel:2014cha}:  the symmetric orbifold theory contains a small ${\cal N}=4$ superconformal ${\cal W}^{s}_\infty$ subalgebra that appears (as a limiting case) in the supersymmetric version of the 3d/2d higher spin/CFT duality \cite{Gaberdiel:2010pz,Gaberdiel:2013vva}. The actual symmetry algebra of the symmetric orbifold theory is however much larger  than this ${\cal W}^{s}_\infty$ algebra: it contains many additional generators that sit in specific representations of the ${\cal W}^{s}_\infty$ algebra. The resulting structure is sometimes referred to as the `Higher Spin Square' \cite{Gaberdiel:2015mra}. 
\smallskip

Given the analogy with free super Yang Mills, one may expect that the symmetric orbifold theory is a very special point in the moduli space of CFT duals, and that it has in fact, at least locally, the largest possible symmetry algebra. It is the aim of this paper to provide evidence for this. More specifically, we shall analyse whether the symmetric orbifold algebra has any deformation parameter, and we shall find that, at least to the level to which we have been able to check this, the algebra is uniquely determined by consistency. 

The argument is slightly subtle and proceeds in two steps. We shall first concentrate on the small ${\cal N}=4$ superconformal ${\cal W}^s_\infty$ subalgebra. There are two natural classes of deformations  we may consider. We may either insist on the deformed algebra to also have small ${\cal N}=4$ superconformal symmetry, or we may allow the deformation to give rise to an algebra with large ${\cal N}=4$ superconformal symmetry.\footnote{We are only interested in deformations that preserve ${\cal N}=4$ superconformal symmetry, i.e.\ either have small ${\cal N}=4$ or large ${\cal N}=4$ superconformal symmetry. The small ${\cal N}=4$ superconformal algebra arises in the degeneration limit of the large ${\cal N}=4$ superconformal algebra.} The first option is what we shall consider in this paper, and the main upshot of our analysis is that no such deformation seems to exist, i.e.\ that up to the level we have analysed the structure, the most general small ${\cal N}=4$ ${\cal W}^s_\infty$ algebra with the appropriate spectrum is uniquely characterised by the central charge without any  additional free parameter. 

This leaves us with the second option, and this is in fact possible \cite{Beccaria:2014jra}: the small ${\cal N}=4$ superconformal ${\cal W}^s_\infty$ algebra is part of a 1-parameter family of large ${\cal N}=4$ ${\cal W}_\infty[\lambda]$ algebras, from which it arises in the limit $\lambda\rightarrow 0$. While it is possible to deform the ${\cal W}^s_\infty$ algebra in this manner, it is easy to see --- this was already clear from the analysis of \cite{Gaberdiel:2014cha} --- that the full symmetry algebra of the symmetric orbifold theory cannot be extended to $\lambda\neq 0$ since the conformal dimensions of the additional generators are not (half)-integer for $\lambda\neq 0$. Combining these two arguments, we conclude that the symmetry algebra of the symmetric orbifold theory does not seem to have any deformation parameter.

We should note that this argument does not rule out the possibility that the small ${\cal N}=4$ ${\cal W}^s_\infty$ algebra may not close on itself in the deformed algebra. Similarly, the structure constants of the additional generators may be deformation dependent (but not those of the small ${\cal N}=4$ ${\cal W}^s_\infty$ subalgebra). Both of these possibilities only affect the structure of OPEs at higher spins, and given that our analysis is based on the structure of the low spin OPEs, we cannot exclude either of these possibilities. However, given the experience with similar situations, in particular, the bosonic and ${\cal N}=2$ superconformal ${\cal W}_{\infty}$ algebra for which the relevant deformation parameters appeared already at very low spins \cite{Gaberdiel:2012ku,Candu:2012tr}, it seems plausible that also in the present context such a deformation (that only affects higher spin OPEs) does not arise.\footnote{We thank Lorenz Eberhardt for a discussion about this point.} 
\medskip

The paper is organised as follows. We begin in Section~\ref{sec:multiplets} with describing the spin spectrum of the small ${\cal N}=4$ ${\cal W}^s_\infty$ algebra we will be studying, and introduce our notation. In Section~\ref{sec:3} we make the most general ansatz for a ${\cal W}^s_\infty$ algebra with this spin spectrum, and analyse the constraints that consistency (i.e.\ the associativity of the OPE, or equivalently the Jacobi identities of the commutators) imposes on the {\it a priori} undetermined OPE coefficients. Our main result is that, to the level we have analysed this problem, there is no free parameter, thus suggesting that the algebra does not have any free parameter. In Section~\ref{sec:4} we subject our analysis to a number of consistency checks; in particular, we compare our results to those of \cite{Beccaria:2014jra}, where a similar analysis for the large ${\cal N}=4$ ${\cal W}_\infty[\lambda]$ algebra was performed, and we study the truncation pattern for the algebra at small values of the central charge. In Section~\ref{sec:5} we conclude with some outlook for future directions. There are also a number of appendices where some of the detailed explicit formulae are given.

\section{The structure of the higher spin multiplets}\label{sec:multiplets}

In this section we set out our notation and explain precisely which algebra we shall be considering in the following.

\subsection{The small ${\cal N}=4$ superconformal algebra}

Let us start by describing the small ${\cal N}=4$ superconformal algebra. In OPE language it is given by 
\cite{Ademolloetalplb,Ademolloetalnpb}, see also \cite[eq.~(9)]{MU} and \cite{Eguchi:1987sm,Eguchi:1987wf}
 \bea
L(z)\, L(w) &\sim&
\frac{1}{(z-w)^4}\, \frac{c}{2}
+\frac{1}{(z-w)^2}\, 2\, L(w)
+\frac{1}{(z-w)}\, \partial L(w)\ ,
\nonu\\
L(z)\, G^{+ a}(w) &\sim&
\frac{1}{(z-w)^2}\, \frac{3}{2}\, G^{+ a}(w)
+\frac{1}{(z-w)}\, \partial G^{+ a}(w)\ ,
\nonu\\
L(z)\, G^{-}_{\,\,a}(w) &\sim&
\frac{1}{(z-w)^2}\, \frac{3}{2}\, G^{-}_{ \,\,a}(w)
+\frac{1}{(z-w)}\, \partial G^{-}_{ \,\,a}(w)\ ,
\eea
\bea
L(z)\, A^{i}(w) &\sim&
\frac{1}{(z-w)^2}\, A^{i}(w)
+\frac{1}{(z-w)}\, \partial A^{i}(w) \ , 
\nonu\\
G^{+a}(z) \, G^{-}_{\,\,b}(w) &\sim& \frac{1}{(z-w)^3}\,\frac{2\,c}{3}\,\delta^{a}\,_{b}
-\frac{4}{(z-w)^2}\, (\sigma^{i})^{a}\,_b \, A^{i}(w) 
\nonu\\
& & +\, 
\frac{1}{(z-w)}\, \Bigg[\,\delta^{a}\,_{b}\,2\,L - 2 (\sigma^{i})^{a}\,_b\,\partial A^{i} \,\Bigg](w)\ ,
\nonu\\
G^{+ a}(z) \, G^{+ b}(w) &\sim & 0 \ ,
\qquad \qquad
G^{-}_{ \,\,a}(z) \, G^{-}_{ \,\,b}(w) \sim  0 \ ,
\nonu\\
A^{i}(z)\,G^{+a}(w)  & \sim &
-\frac{1}{(z-w)}\, \frac{1}{2}\,(\sigma^{i})^{a}\,_{b} \, G^{+ b}(w)\ ,
\eea
\bea
A^{i}(z)\,G^{-}_{\,\,a}(w)  & \sim &
\frac{1}{(z-w)}\, \frac{1}{2}\,G^{-}_{ \,\,b}
\, (\sigma^{i})^{b}\,_{a} (w)
\ , 
\nonu\\
A^{i}(z)A^{j}(w) & \sim &
\frac{1}{(z-w)^2}\,\frac{c}{12} \, \delta^{ij}
+\frac{1}{(z-w)}\,i\,\varepsilon_{ijk}\,A^{k}(w)\ .
\label{OPELGA}
\eea
Here, the three spin-$1$ currents $A^i$, $i=1,2,3$ generate the affine Kac-Moody algebra $\mathfrak{su}(2)$ at level $k=\frac{c}{6}$, where $c$ is the central charge of the Virasoro algebra. The four supercharge currents $G^{+ a}(z)$ and $G^{-}_{\,\,a}(z)$, $a=1,2$, sit in a doublet with respect to the $\mathfrak{su}(2)_R$ algebra that is generated by the zero modes $A^i_0$. The matrices $(\si_i)^{a}\,_{b}$ are the Pauli matrices ($i=1,2,3$), satisfying $\sigma^i \cdot \sigma^j = \delta^{ij} \,{\bf 1} + i \epsilon_{ijk} \sigma^k$, where $\varepsilon_{ijk}$ is the totally antisymmetric tensor with $\varepsilon_{123}=1$. Finally, the antisymmetric tensors $\varepsilon_{ab}$ and
$\varepsilon^{ab}\,(=-\varepsilon_{ab})$ with $\varepsilon_{12}=1$ can be used to raise and
lower the $\mathfrak{su}(2)_R$ spinor indices $a, b, \ldots$, while the $\mathfrak{su}(2)$ indices $i,j,$ are raised and lowered with $\delta_{ij}=\delta^{ij}$. Repeated indices are summed over.\footnote{Since raising or lowering the $i,j$ indices is trivial, there is no difference whether these indices are upstairs or downstairs.}

\subsection{The spectrum of the algebra}

As was explained in the introduction, the symmetric orbifold theory contains the small ${\cal N}=4$ 
${\cal W}^s_{\infty}$ algebra that arises in the limit $\lambda\rightarrow 0$ from the large ${\cal N}=4$  algebras ${\cal W}_{\infty}[\lambda]$. The large ${\cal N}=4$ superconformal algebra contains two $\mathfrak{su}(2)$ algebras (at levels $k_+$ and $k_-$), and the  ${\cal W}_\infty[\lambda]$ algebra is generated (in addition to the large ${\cal N}=4$ superconformal algebra) by the spin fields from the multiplets  
\be
\begin{array}{llc}\label{D2mult2}
& s: & ({\bf 1},{\bf 1})  \\
& s+\tfrac{1}{2}: & ({\bf 2},{\bf 2})  \\
R^{(s)}: \qquad & s+1: & ({\bf 3},{\bf 1}) \oplus ({\bf 1},{\bf 3})  \\
& s+\tfrac{3}{2}: & ({\bf 2},{\bf 2})  \\
& s+2: & ({\bf 1},{\bf 1}) 
\end{array}
\ee
where the two bold-face numbers describe the representations with respect to the two $\mathfrak{su}(2)$ algebras, and $s=1,2,\ldots$. The parameter $\lambda$ is defined to be $\lambda = \frac{k_-}{k_+ + k_- + 2}$. 

The limit $\lambda\rightarrow 0$ is obtained by taking $k_+\rightarrow \infty$. In this limit the first $\mathfrak{su}(2)$ algebra decouples, and we should only organise the states with respect to the second algebra. This is to say, the spin spectrum of the small ${\cal N}=4$ algebra we are interested in has the form 
\be
\begin{array}{llc}\label{smallN4}
& s: & {\bf 1}  \\
& s+\tfrac{1}{2}: & 2 \cdot {\bf 2}  \\
R^{(s)}: \qquad & s+1: & {\bf 3} \oplus 3\cdot {\bf 1}  \\
& s+\tfrac{3}{2}: & 2 \cdot {\bf 2}  \\
& s+2: & {\bf 1} 
\end{array}
\ee
We will use the convention that the various components of this multiplet are denoted as described in Table~1. 

\begin{table}[ht]\label{Tab01}
\rule{0pt}{8ex}
\centering 
\begin{tabular}{|c|c|c|c|c|c| } 
\hline 
$\mbox{components }$ &   \T $\Phi^{(s)}_0$ \B & 
\T $\Phi^{(s),+ a}_{\frac{1}{2}}\  \Phi^{(s),-}_{\frac{1}{2} \,\,\,\,\,\,\,a}$ \B
& \T $\Phi^{(s),i}_1  \  \Phi^{(s)}_1 \ \Phi^{(s),+ +}_1 \   \Phi^{(s),- -}_1$ \B 
& \T  $\Phi^{(s),+ a}_{\frac{3}{2}}\ \Phi^{(s),-}_{\frac{3}{2} \,\,\,\,\,\,\, a}$  \B &
\T $\Phi^{(s)}_2 $ \B \\[1.5ex] 
\hline
\T \T \mbox{conformal spin} \B & $s$ & $s+\frac{1}{2}$ & $s+1$ &
$s+\frac{3}{2}$ & $s+2$ \\ 
[1.5ex]
\hline
 \T \mbox{$\mathfrak{su}(2)$  rep. }& ${\bf 1}$ &
  ${\bf 2} \oplus {\bf 2}$ &
  ${\bf 3} \oplus {\bf 1} \oplus {\bf 1} \oplus {\bf 1}$
  & ${\bf 2} \oplus {\bf 2}$
  & ${\bf 1}$ \\
[1.5ex]
\hline
\end{tabular}
\caption{The conformal spins and $\mathfrak{su}(2)$
  representations of the ${\cal N}=4$ multiplet.}
\end{table}

We will work with the convention that all of these fields (with the exception of $\Phi^{(s)}_2$, see below) are primary fields with respect to the stress energy tensor. This is to say, they satisfy (for $n<2$)
\be
L(z)\, \Phi^{(s)*}_n (w) \sim \frac{(s+n)}{(z-w)^2} \, \Phi^{(s)*}_n (w) + \frac{1}{(z-w)}\,  \partial \Phi^{(s)*}_n (w) \ , 
\ee
where $*$ denotes any of the additional indices. On the other hand, for the last component we make the ansatz that it is only quasiprimary and satisfies
\be
L(z)\, \Phi^{(s)}_2(w) \sim
\frac{t(s)}{(z-w)^4}\, \Phi^{(s)}_0(w)
+\frac{(s+2)}{(z-w)^2}\, \Phi^{(s)}_2(w)
+\frac{1}{(z-w)}\, \partial \Phi^{(s)}_2(w)\ ,
\ee
where $t(s)$ is fixed by the Jacobi identities, and turns out to equal (see also \cite[eq.~(2.12)]{Beccaria:2014jra}) 
\be
t(s) =-\frac{12\,s(s+1)}{(2s+1)}\ .
\ee
Here the Jacobi identities are those that arise from considering two fields from the small ${\cal N}=4$ algebra together with one field from the spin-$s$ multiplet.
These Jacobi identities also allow us to fix the $s$-dependent coefficients in the OPEs with the $\mathfrak{su}(2)$ currents, and a consistent convention is 
\bea
A^{i}(z) \, \Phi^{(s)}_0(w) 
&\sim&
0 \ ,
\nonu\\
A^{i}(z) \, \Phi^{(s),+a}_\frac{1}{2}(w) 
&\sim&
- \frac{1}{(z-w)}\, \frac{1}{2} \,(\sigma^{i})^{a}\,_{b} \,\Phi^{(s),+ b}_\frac{1}{2}(w)\ ,
\nonu\\
A^{i}(z) \,\Phi^{(s),-}_{\frac{1}{2}\,\,\,\,\,\,\,a}(w) 
&\sim&
- \frac{1}{(z-w)}\, \frac{1}{2} \,(\sigma^{i})_{a}\,^{b}(w)  \,\Phi^{(s),-}_{\frac{1}{2}\,\,\,\,\,\,\,b}
\ ,
\nonu\\
A^{i}(z) \, \Phi^{(s)}_1(w) &\sim& 0\ ,
\nonu\\
A^{i}(z) \, \Phi^{(s),\pm \pm}_1(w) &\sim&0\ ,
\nonu\\
A^{i}(z) \, \Phi^{(s),j}_1(w) &\sim&
\frac{1}{(z-w)^2}\,s\,\delta^{ij}\,\Phi^{(s)}_0(w)
+\frac{1}{(z-w)}\,i\,\varepsilon_{ijk} \Phi^{(s),k}_1(w)\ ,
\eea
\bea
A^{i}(z) \, \Phi^{(s),+ a}_\frac{3}{2}(w) & \sim &
-\frac{1}{(z-w)^2}\,\frac{2s(s+1)}{(2s+1)}\,(\sigma^{i})^{a}\,_{b}\, \Phi^{(s),+ b}_\frac{1}{2}(w)
-\frac{1}{(z-w)}\,\frac{1}{2}\,(\sigma^{i})^{a}\,_{b} \,\Phi^{(s),+ b}_\frac{3}{2}(w)\ ,
\nonu\\
A^{i}(z) \, \Phi^{(s),-}_{\frac{3}{2}\,\,\,\,\,\,\,a}(w) & \sim &
-\frac{1}{(z-w)^2}\,\frac{2s(s+1)}{(2s+1)}\,(\sigma^{i})_{a}\,^{b}\, \Phi^{(s),-}_{\frac{1}{2}\,\,\,\,\,\,\,b}
(w) 
-\frac{1}{(z-w)}\,\frac{1}{2}\,(\sigma^{i})_{a}\,^{b}\, \Phi^{(s),-}_{\frac{3}{2}\,\,\,\,\,\,\,b}(w)
\ ,
\nonu\\
A^{i}(z) \, \Phi^{(s)}_2(w) & \sim &
-\frac{1}{(z-w)^2}\,2(s+1)\,\Phi^{(s),i}_1 (w)\ .
\label{OPEAandother}
\eea
Similarly, we can determine the OPEs with the supercurrents, and the explicit formulae are given in Appendix~\ref{app:OPEsuper}.

\section{The structure of the small ${\cal N}=4$ ${\cal W}^s_\infty$ algebra}\label{sec:3}

Having fixed these conventions we can now proceed to study the most general ${\cal W}$ algebra with this spin spectrum. The method we shall use is similar to what was done in \cite{Beccaria:2014jra}. However, since the constraints that come from the $\mathfrak{su}(2)$ symmetry are now weaker than for the case that was studied there --- compare (\ref{smallN4}) and (\ref{D2mult2}) --- we cannot directly use the results from  \cite{Beccaria:2014jra}, but rather have to start from scratch again. 


In order to introduce a convenient notation, we write the OPE of two quasiprimary fields of spins $h_i$ and $h_j$ as \cite{Blumenhagen:1990jv}
\bea
\phi^i(z)\, \phi^j(w)= \sum_{k}  \frac{{C^{ij}}_k}{(z-w)^{h_i+h_j-h_k}}  \sum_{n=0}^{\infty}  
\frac{(h_i-h_j+h_k)_n}{n!(2h_k)_n}\, (z-w)^n \, \partial^n
\phi^k(w) \ ,
\label{eq:opefull}
\eea
where $k$ labels the different quasiprimary fields that appear on the right-hand-side, and $(x)_n$ stands for the Pochhammer symbol. We shall also sometimes use the shorthand notation for the singular part of the OPE (\ref{eq:opefull}) as 
\begin{equation}
\phi^i \times \phi^j \sim \sum_{k\, :\, h_k < h_i+h_j}  {C^{ij}}_k \, \phi^k\ .
\label{eq:opeshort}
\end{equation}
The basic strategy is to make the most general ansatz for the various OPEs, and then to determine the structure constants $ {C^{ij}}_k$ recursively, using the various Jacobi identities. The quasiprimary fields that appear on the right-hand-side may either be fundamental or composite fields; in order to write down the most general ansatz, it is therefore important to enumerate all the different composite fields that may appear. 

\subsection{Composite fields}

A convenient way of enumerating all the composite fields is by using character techniques, as was also done in \cite{Beccaria:2014jra}. To this end we introduce a `mark' for each fundamental field of the algebra; for the superconformal generators this is given in Table~2, while the marks for the other fields are specified in Table~3.

\begin{table}[ht]
\centering 
\begin{tabular}{|c|c|c|c|c|c| } 
\hline 
$\T \mbox{component}$ \B & $A^{ i} $ & $ G^{+ a} $ & $ G^{-}_{\,\,a} $ &
$L $\\ \hline
\T $\mbox{mark}$ \B & $y_{0,1} $ & $y^{+}_{0,3/2}$ & $y^{-}_{0,3/2}$ &
$ y_{0,2} $ \\ 
[1.5ex]
\hline
\end{tabular}
\caption{The marks for the generators of
  the small ${\cal N}=4$ superconformal algebra.
} 
\end{table}
\begin{table}[ht]
\centering 
\begin{tabular}{|c|c|c|c|c|c|c|c|c|c| } 
\hline 
\T $\mbox{component}$ \B & $\Phi^{(s)}_0$ & $\Phi^{(s),+ a}_{1/2}$ &
$\Phi^{(s),-}_{1/2\,\,\,\,\,\,\,a}$ & $\Phi^{(s)}_{1}$ &
$\Phi^{(s), \pm \pm}_{1}$ & $\Phi^{(s), i}_{1}$ &
$\Phi^{(s),+ a}_{3/2} $ & $\Phi^{(s),-}_{3/2\,\,\,\,\,\,\,a}$ &
$\Phi^{(s)}_2$ \\ \hline
\T $\mbox{mark}$ \B &
$ y_{s,0}$ & $y^{+}_{s,1/2}$ & $y^{-}_{s,1/2}$ & $y^*_{s,1}$ &
$y^{\pm \pm}_{s,1}$ & $y_{s,1}$ & $y^{+}_{s,3/2}$ & $y^{-}_{s,3/2}$ &
$y_{s,2}$ \\ 
[1.5ex]
\hline
\end{tabular}
\caption{The marks for the higher spin operators.
} 
\end{table}
\smallskip

\noindent Then we consider the `marked' vacuum character of the full small ${\cal N}=4$ ${\cal W}^s_\infty$ algebra
\be
\chi_{\infty}= \chi_0 \cdot \chi_{\rm hs} \ , 
\label{chiinfty}
\ee
where $\chi_0$ is the vacuum character of the small ${\cal N}=4$ algebra, see e.g.\ \cite{Eguchi:1987sm,Eguchi:1987wf}
\bea
\chi_0 = \prod_{n=1}^\infty \frac{\prod^{\frac{1}{2}}_{m=-\frac{1}{2}}(1+y^+_{0,\frac{3}{2}}\,z^{2m}\,q^{n+\frac{1}{2}})(1+y^-_{0,\frac{3}{2}}\,z^{2m}\,q^{n+\frac{1}{2}})}
{(1-y_{0,2}\,q^{n+1})
\prod^1_{m=-1}(1-y_{0,1}\,z^{2m}\,q^n)} \ ,
\label{chizero}
\eea
while $\chi_{\rm {hs}}$ is the contribution from the higher spin fields 
\bea
\chi_{\rm {hs}}=  \prod_{s=1}^\infty\prod_{n=s}^\infty \frac{\prod^{\frac{1}{2}}_{m=-\frac{1}{2}}\prod_{a\in \{+,-\}}(1+y^a_{s,\frac{1}{2}}z^{2m}\, q^{n+\frac{1}{2}}) (1+y^a_{s,\frac{3}{2}}\, z^{2m}\ q^{n+\frac{3}{2}})
}
{(1-y_{s,0}q^n)\prod_{b\in \{\,*,\pm\pm\}}(1-y^b_{s,1}q^{n+1})\prod^1_{m=-1}(1-y_{s,1}z^{2m} q^{n+1}) (1-y_{s,2}q^{n+2})}\ .
\nonu
\eea
Here we have also denoted by $z$ the chemical potential for the $\mathfrak{su}(2)$
algebra, while $q$ is, as usual, counting the $L_0$ eigenvalues.  
In order to read off the quasiprimary fields, we now rewrite $\chi_\infty$ in terms of $\mathfrak{sl}(2,\mathbb{R})$ representations as 
\bea
\chi_\infty = 1+ \sum_{s\in\mathbb{N}/2}\frac{d_s \, q^s}{(1-q)}\ .
\eea
The numbers $d_s$ then count the number of quasiprimary fields at spin $s$. Up to $s=3$ they are explicitly given as 
\bea
d_1 & =  &y_{0,1}\mbox{ch}_1(z)+y_{1,0}\ ,
\nonu\\
d_{\frac{3}{2}} & =& \big(y^+_{0, \frac{3}{2}}+y^-_{0, \frac{3}{2}}+y^+_{1, \frac{1}{2}}+y^-_{1, \frac{1}{2}}\big)\mbox{ch}_\frac{1}{2}(z)\ ,
\nonu\\
d_{2} & =& y_{0,2}+y_{2,0}+y^*_{1,1}+y^{++}_{1,1}+y^{--}_{1,1}+(y_{1,0})^2+(y_{0,1})^2 (\mbox{ch}_2(z)+1)+y_{0,1}y_{1,0}\mbox{ch}_1(z)
\nonu \\
&   & + \, y_{1,1}\mbox{ch}_1(z)\ ,
\nonu\\
d_{\frac{5}{2}} &=& \big( y^+_{2, \frac{1}{2}}+ y^-_{2, \frac{1}{2}}+y^+_{1, \frac{3}{2}}+y^-_{1, \frac{3}{2}}\big)\mbox{ch}_\frac{1}{2}(z) \nonu \\
& & + \, 
\big( y^+_{1, \frac{1}{2}}+y^-_{1, \frac{1}{2}}+y^+_{0, \frac{3}{2}}+y^-_{0, \frac{3}{2}}\big)
\big( y_{1,0}+y_{0,1}\mbox{ch}_1(z) \big)\mbox{ch}_\frac{1}{2}(z)\ ,
\nonu\\
d_{3} &=&y_{3,0}+y_{1,2}+y^*_{2,1}+y^{++}_{2,1}+y^{--}_{2,1}+y_{2,1}\mbox{ch}_1(z)
+y_{0,1}(y_{1,0})^2 \mbox{ch}_1(z)
\nonu\\
& & + \, (y_{0,1})^2 y_{1,0}\big(\mbox{ch}_2(z)+1\big) + 
\big(y_{1,0}+y_{0,1}\mbox{ch}_1(z)\big)\big( y^*_{1,1}+y^{++}_{1,1}+y^{--}_{1,1}+y_{0,2}+y_{2,0} \big)
\nonu\\
& & +\, (y_{1,0})^3+(y_{0,1})^3\big(\mbox{ch}_3(z)+\mbox{ch}_1(z)\big)
+y_{0,1}y_{1,1}\big(\mbox{ch}_2(z)+\mbox{ch}_1(z)+1\big)
+y_{1,0}y_{1,1}\mbox{ch}_1(z)
\nonu\\
& &  \, 
+ \big(y^+_{0, \frac{3}{2}} y^-_{0, \frac{3}{2}}+y^+_{0, \frac{3}{2}} y^+_{1, \frac{1}{2}} +y^+_{0, \frac{3}{2}} y^-_{1, \frac{1}{2}} 
+y^-_{0, \frac{3}{2}} y^+_{1, \frac{1}{2}} +y^-_{0, \frac{3}{2}} y^-_{1, \frac{1}{2}} 
+y^+_{1, \frac{1}{2}} y^-_{1, \frac{1}{2}}  \big) (\mbox{ch}_1(z)+1)
\nonu\\
& & + \,  \big(y_{0,1}y_{1,0}+(y_{0,1})^2\big)\mbox{ch}_1(z) + (y^+_{0, \frac{3}{2}})^2+(y^-_{0, \frac{3}{2}})^2
+ (y^+_{1, \frac{1}{2}})^2+(y^-_{1, \frac{1}{2}})^2
\ ,
\eea
where $\mbox{ch}_j(z) \equiv \sum_{m=-j}^j \, z^{2m}$
is the character of the $\mathfrak{su}(2)$ representation
of spin $j$.\footnote{These results agree, upon setting $z_+=1$ and $z_-=z$ with the corresponding formulae in \cite{Beccaria:2014jra} once some typos there have been corrected.} The corresponding quasiprimary fields can then be constructed using the quasiprimary normal ordered product $[\phi^i\phi^j]$, which is defined as \cite{Blumenhagen:1990jv}
\bea\label{nop}
  (\phi^i\phi^j) = [\phi^i\phi^j]  + \sum_k \, {C^{ij}}_k
\left(
\begin{array}{c} 
  2h_i-1 \\
  h_i+h_j-h_k
\end{array}
 \right)
\frac{\Gamma(2h_k)}{\Gamma(h_i+h_j+h_k)}\, \partial^{h_i+h_j-h_k}\phi^k \ . 
\eea
For example, for the case of the stress-energy tensor, $[L L] = (L L)-\frac{3}{10} \pa^2 L$, leading to the well-known quasiprimary field $L L -\frac{3}{10} \pa^2 L$. 

Using the convention that $[\phi^i \phi^j \phi^k]= [\phi^i [\phi^j \phi^k]]$, the quasiprimary fields up to spin $s=3$ are then explicitly given as 
\bea
s = 1&:&\quad
\ A^{ i}\ , \ \Phi^{(1)}_0 \ ,
\nonu \\
s= \frac{3}{2}&:&\quad 
G^{+ a}\ ,\ G^{-}_{\,\,a}\ ,\ \Phi^{(1),+ a}_{\frac{1}{2}}\ ,\ \Phi^{(1),-}_{\frac{1}{2}\,\,\,\,\,\,\,a}\ ,
\nonu \\
s= 2&:&\quad 
L\ ,  
\ \Phi^{(2)}_0 \ , 
\ \Phi^{(1)}_1\  , 
\ \Phi^{(1),\pm \pm}_1  , \ \Phi^{(1),i}_1
\ ,  \ [A^i A^j] \ ,  \ [A^i  \Phi^{(1)}_0] \ , \ [\Phi^{(1)}_0 \Phi^{(1)}_0] \ , 
\nonu \\
s= \frac{5}{2}&:&\quad 
\Phi^{(2)+ a}_{\frac{1}{2}}\ ,
\Phi^{(2),-}_{\frac{1}{2}\,\,\,\,\,\,\,a}\ ,
\ \Phi^{(1),+ a}_{\frac{3}{2}}\ ,
\ \Phi^{(1),-}_{\frac{3}{2}\,\,\,\,\,\,\,a}\ ,
\ [\Phi^{(1)}_0 \Phi^{(1),+ a}_{\frac{1}{2}}]\ ,
\ [\Phi^{(1)}_0 \Phi^{(1),-}_{\frac{1}{2}\,\,\,\,\,\,\,a}]\ ,
\nonu\\
&& \quad
\ [G^{+ a} \Phi^{(1)}_0 ]\ ,
\ [G^{-}_{\,\,a} \Phi^{(1)}_0 ]\ ,
\ [G^{+ a} A^i ]\ ,
\ [G^{-}_{\,\,a} A^i ]\ ,
\ [A^i \Phi^{(1),+ a}_{\frac{1}{2}} ]\ ,
\ [A^i \Phi^{(1),-}_{\frac{1}{2}\,\,\,\,\,\,\,a}\, ]\ ,
\eea
\bea
s = 3&:&\quad
\Phi^{(3)}_0\ , 
\ \Phi^{(2)}_1\ , 
\ \Phi^{(2),\pm \pm}_{1}\ , 
\ \Phi^{(2),i}_{1}\ , 
\ \Phi^{(1)}_{2}\ , 
\ [\Phi^{(1)}_0 \Phi^{(1)}_0\Phi^{(1)}_0]\ , 
\ [\Phi^{(1)}_0 \Phi^{(2)}_0]\ , 
\ [\Phi^{(1)}_0 \Phi^{(1)}_1]\ , 
\nonu\\
&& \quad
[\Phi^{(1)}_0 \Phi^{(1),\pm \pm}_1]\ , 
\ [\Phi^{(1)}_0 \Phi^{(1),i}_1]\ , 
\ [A^i\Phi^{(2)}_0]\ ,
\ [A^i\Phi^{(1)}_1]\ ,
\ [A^i\Phi^{(1),\pm \pm}_1]\ ,
\ [A^i\Phi^{(1),j}_1]\ ,
\ [L A^i ]\ ,
\nonu\\
& &
\quad
 [L \Phi^{(1)}_0]\ ,
\ [A^i A^j A^k ]\ ,
\ [A^i A^j \Phi^{(1)}_0 ]\ ,
\ [A^i  \Phi^{(1)}_0 \Phi^{(1)}_0 ]\ ,
\ [A^i A^j]_{-1}\ ,
\ [A^i \Phi^{(1)}_0 ]_{-1}\ ,
\nonu\\
& &
\quad
[\Phi^{(1),+ a}_\frac{1}{2} \Phi^{(1),-}_{\frac{1}{2}\,\,\,\,\,\,\,b}\, ]\ , 
\varepsilon_{ab } \ [\Phi^{(1),+a}_\frac{1}{2} \Phi^{(1),+ b}_{\frac{1}{2}}]\ , 
\varepsilon^{ab } \ [\Phi^{(1),-}_{\frac{1}{2}\,\,\,\,\,\,\,a} \Phi^{(1),-}_{\frac{1}{2}\,\,\,\,\,\,\,b}\, ]\ , 
\ [G^{+ a} G^{-}_{\,\,b}]\ ,
\varepsilon_{ab } \ [G^{+ a} G^{+ b}]\ ,
\nonu\\
& &
\quad
\varepsilon^{ab }\ [G^{-}_{\,\,a} G^{-}_{\,\,b}]\ ,
\ [G^{+ a} \Phi^{(1),+ b}_\frac{1}{2}]\ ,
\ [G^{+ a} \Phi^{(1),-}_{\frac{1}{2}\,\,\,\,\,\,\,b}\, ]\ ,
\ [G^{-}_{\,\,a} \Phi^{(1),+ b}_\frac{1}{2}]\ ,
\ [G^{-}_{\,\,a} \Phi^{(1),-}_{\frac{1}{2}\,\,\,\,\,\,\,b}\, ]\ .
\eea
Here we have used the short-hand notation 
\bea
[A^i \Phi^{(s)}_0]_{-1}=\frac{1}{2}\,(\partial A^i   \Phi^{(s)}_0 )-\frac{1}{2}\,( A^i   \partial  \Phi^{(s)}_0 )\ ,\qquad
[A^{i} A^{j}]_{-1}=-\frac{1}{2}\,(\partial A^i  A^j)+\frac{1}{2}\,( A^i  \partial  A^j)\ .
\eea

\subsection{Ansatz for OPEs}

With these preparations we can now make the most general ansatz for the OPEs of the (low-lying) higher spin fields. (Recall that the OPEs involving at least one ${\cal N}=4$ superconformal field have already been fixed above.) 
 We shall then recursively impose the Jacobi identities to constrain the coefficients with which the (quasi-primary) fields appear on the right-hand-side of the various OPEs. One direct consequence of these Jacobi identities is that the 
$\mathfrak{su}(2)$ symmetry coming from the zero modes of the $A^i$ currents must be respected by the OPE, and we shall therefore implement this constraint already in our ansatz. We shall organise the OPEs by their `total  spin', i.e.\ by the sum of the spins of the two fields whose OPE we consider. The following calculations were performed with the help of the {\tt Mathematica} package of Thielemans \cite{T}. 

\subsubsection{The case of total spin $2$}

At total spin $2$ the only OPE to consider is 
\bea
\Phi^{(1)}_0\times\Phi^{(1)}_0  \sim 
n_1\, I +  0 \cdot \Phi^{(1)}_0\ .
\label{Ansatz2}
\eea
The right hand side can have at most conformal weight $1$, and the $\mathfrak{su}(2)$ symmetry implies that only the identity $I$, or $\Phi^{(1)}_0$ itself can appear. Finally, we use that the coefficient in front of $\Phi^{(1)}_0$ must be zero by the symmetry of the $3$-point function $\langle \Phi^{(1)}_0 \, \Phi^{(1)}_0 \, \Phi^{(1)}_0 \rangle$.\footnote{Since $J\equiv \Phi^{(1)}_0$ is a spin-$1$ field, we can think of the OPE as defining an affine Kac-Moody algebra. Because of the anti-symmetry of the bracket the term in the commutator proportional to $[J_m,J_n] \sim f J_{m+n}$ must vanish, i.e.\ $f=0$.} The coefficient $n_1$ is a normalisation constant, and we shall keep it arbitrary for the time being.

\subsubsection{The case of total spin $\frac{5}{2}$}

In this case, the only OPEs to consider are 
\bea
\Phi^{(1)}_0 \times \Phi^{(1),+ a}_\frac{1}{2}& \sim &
w_1\,G^{+a} +0\cdot  \, \Phi^{(1),+a}_\frac{1}{2}  \,,
\qquad
\Phi^{(1)}_0 \times \Phi^{(1),-}_{\frac{1}{2}\,\,\,\,\,\,\,a} \sim 
w_2\,G^{-}_{\,\, a}   +0\cdot  \, \Phi^{(1),-}_{\frac{1}{2}\,\,\,\,\,\,\,a} \ , 
\label{Ansatz5H}
\eea
where we have used the $\mathfrak{su}(2)$ symmetry that implies that the right-hand-side must be a doublet.  The coefficients in front of $\Phi^{(1),+ a}_\frac{1}{2}$ and $\Phi^{(1),-}_{\frac{1}{2}\,\,\,\,\,\,\,a}$ 
have to vanish because the $3$-point function
\be
\braket{\Phi^{(1)}_0(z_1)   \Phi^{(1)}_\frac{1}{2}(z_2)  \Phi^{(1)}_\frac{1}{2}(z_3)}  
\ee
must be anti-symmetric under the exchange of $z_2$ and $z_3$ (since $\Phi^{(1)}_\frac{1}{2}$ is a fermionic field).

\subsubsection{The case of total spin $3$ }

At total spin-$3$, the most general ansatz (respecting the $\mathfrak{su}(2)$ symmetry) is 
\bea
\Phi^{(1)}_0 \times \Phi^{(1)}_1 
&\sim& 
w_3\, \Phi^{(1)}_0
+w_4\, L
+w_5\, [A^i A^i]
+w_6\, [\Phi^{(1)}_0  \Phi^{(1)}_0]
+w_7\, \Phi^{(2)}_0
+w_{8}\, \Phi^{(1)}_1
\ ,
\nonu\\
\Phi^{(1)}_0 \times \Phi^{(1),++}_1 
&\sim&
w_{9}\, \Phi^{(1),++}_1
\ ,
\qquad\qquad 
\Phi^{(1)}_0 \times\Phi^{(1),--}_1 
\sim
w_{10}\, \Phi^{(1),--}_1
\ ,
\nonu\\
\Phi^{(1)}_0 \times \Phi^{(2)}_0
&\sim&
w_{11}\, \Phi^{(1)}_0
+w_{12}\, L
+w_{13}\, [A^i A^i]
+w_{14}\, [\Phi^{(1)}_0  \Phi^{(1)}_0]
\nonu \\
&  &  +\, w_{15}\, \Phi^{(2)}_0
+w_{16}\, \Phi^{(1)}_1\ ,
\nonu \\
\Phi^{(1)}_0 \times \Phi^{(1),i}_1 
&\sim&
w_{17}\,A^i
+w_{18}\,[A^i \Phi^{(1)}_0 ]+w_{19}\,\Phi^{(1),i}_1 \ ,
\nonu \\
\Phi^{(1),+a}_\frac{1}{2} \times \Phi^{(1),+b}_\frac{1}{2}
&\sim&
\varepsilon^{ab} \,
w_{20}\, \Phi^{(1),++}_1
\ ,
\qquad
\Phi^{(1),-}_{\frac{1}{2}\,\,\,\,\,\,\,a} \times \Phi^{(1),-}_{\frac{1}{2}\,\,\,\,\,\,\,b}
\sim
\varepsilon_{ab} \,
w_{21}\, \Phi^{(1),--}_1
\ ,
\nonu\\
\Phi^{(1),+a}_\frac{1}{2} \times  \Phi^{(1),-}_{\frac{1}{2}\,\,\,\,\,\,\,b}
&\sim&
\delta^{a}\,_{b}\Big(\,
w_{22}\,I +w_{23}\, \Phi^{(1)}_0+w_{24}\,L +w_{25}\,[A^i A^i] +w_{26}\, [\Phi^{(1)}_0 \Phi^{(1)}_0]+w_{27}\,\Phi^{(2)}_0 \nonu\\
&
& \qquad + w_{28}\, \Phi^{(1)}_1
\,\Big)
+(\sigma^{i})^{a}\,_{b}
\Big(\,
w_{29}\,A^i+w_{30}\,[A^i \Phi^{(1)}_0]+w_{31}\,\Phi^{(1),i}_1
\,\Big)\ ,
\label{Ansatz3}
\eea
where we have introduced free parameters $w_i$ that will be subsequently fixed by imposing the Jacobi identities. 

\subsubsection{The case of total spin $\frac{7}{2}$}

The most general ansatz for this case can be described as
\bea
\Phi^{(1)}_0 \times \Phi^{(1),+a}_\frac{3}{2}
&\sim&
w_{32}\,G^{+a}
+w_{33}\,\Phi^{(1),+a}_\frac{1}{2}
+w_{34}\,\Phi^{(2),+a}_\frac{1}{2}
+w_{35}\,\Phi^{(1),+a}_\frac{3}{2}
+w_{36}\,[\Phi^{(1)}_0 \Phi^{(1),+a}_\frac{1}{2}]
\nonu\\
& &  + \, w_{37}\,[G^{+a} \Phi^{(1)}_0]
+
(\sigma^{i})^{a}\,_b
\Big(\,
w_{38}\,[A^i \Phi^{(1),+b}_\frac{1}{2}]
+w_{39}\,[G^{+b} A^i]\,\Big)\ ,
\nonu\\
\Phi^{(1)}_0 \times \Phi^{(1),-}_{\frac{3}{2}\,\,\,\,\,\,\,a}
&\sim&
w_{40}\,G^{-}_{\,\,a}
+w_{41}\,\Phi^{(1),-}_{\frac{1}{2}\,\,\,\,\,\,\,a}
+w_{42}\,\Phi^{(2),-}_{\frac{1}{2}\,\,\,\,\,\,\,a}
+w_{43}\,\Phi^{(1),-}_{\frac{3}{2}\,\,\,\,\,\,\,a}
+w_{44}\,[\Phi^{(1)}_0 \Phi^{(1),-}_{\frac{1}{2}\,\,\,\,\,\,\,a}]
\nonu\\
& & +\, w_{45}\,[G^{-}_{\,\,a} \Phi^{(1)}_0]
+ (\sigma^{i})^{b}\,_a
\,\Big(\,
w_{46}\,[A^i \Phi^{(1),-}_{\frac{1}{2}\,\,\,\,\,\,\,b}]
+w_{47}\,[G^{-}_{\,\,b} A^i]\, \Big)\ ,
\nonu\\
\Phi^{(1)}_0 \times\Phi^{(2),+a}_{\frac{1}{2}}
&\sim&
w_{48}\,G^{+a}+\cdots
+
(\sigma^{i})^{a}\,_b
\Big(\,
w_{54}\,[A^i  \Phi^{(1),+b}_\frac{1}{2}]
+w_{55}\,[G^{+b} A^i]\,\Big) \ ,
\nonu\\
\Phi^{(1)}_0 \times \Phi^{(2),-}_{\frac{1}{2}\,\,\,\,\,\,\,a}
&\sim&
w_{56}\,G^{-}_{\,\,a}+\cdots
+ (\sigma^{i})^{b}\,_a \Big(\,
w_{62}\,[A^i \, \Phi^{(1),-}_{\frac{1}{2}\,\,\,\,\,\,\,b}]
+w_{63}\,[G^{-}_{\,\,b} A^i]\, \Big) \ ,
\nonu\\
\Phi^{(1),+a}_\frac{1}{2} \times \Phi^{(2)}_0
&\sim&
w_{64}\,G^{+a}+\cdots
+
(\sigma^{i})^{a}\,_b
\Big(\,
w_{70}\,[A^i\, \Phi^{(1),+b}_\frac{1}{2}]
+w_{71}\,[G^{+b} A^i]\,\Big) \ ,
\nonu\\
\Phi^{(1),-}_{\frac{1}{2}\,\,\,\,\,\,\,a} \times \Phi^{(2)}_0
&\sim&
w_{72}\,G^{-}_{\,\,a}+\cdots
+ (\sigma^{i})^{b}\,_a \Big(\,
w_{78}\,[A^i\,  \Phi^{(1),-}_{\frac{1}{2}\,\,\,\,\,\,\,b}]
+w_{79}\,[G^{-}_{\,\,b} A^i]\, \Big) \ ,
\nonu\\
\Phi^{(1),+a}_\frac{1}{2} \times \Phi^{(1)}_1
&\sim&
w_{80}\,G^{+a}+\cdots
+
(\sigma^{i})^{a}\,_b
\Big(\,
w_{86}\,[A^i\,  \Phi^{(1),+b}_\frac{1}{2}]
+w_{87}\,[G^{+b} A^i]\,\Big) \ ,
\nonu\\
\Phi^{(1),+a}_\frac{1}{2} \times \Phi^{(1),++}_1
&\sim& 0 \ ,
\nonu\\
\Phi^{(1),+a}_\frac{1}{2}\times \Phi^{(1),--}_1
&\sim&
\varepsilon^{ab}\Big(
\,w_{88}\,G^{-}_{\,\,b}
+w_{89}\,\Phi^{(1),-}_{\frac{1}{2}\,\,\,\,\,\,\,b}
+w_{90}\,\Phi^{(2),-}_{\frac{1}{2}\,\,\,\,\,\,\,b}
+w_{91}\,\Phi^{(1),-}_{\frac{3}{2}\,\,\,\,\,\,\,b}
+w_{92}\,[\Phi^{(1)}_0 \Phi^{(1),-}_{\frac{1}{2}\,\,\,\,\,\,\,b}]
\nonu\\
&&+w_{93}\,[G^{-}_{\,\,b}  \Phi^{(1)}_0]
\,\Big)
+
(\sigma^{i})^{a}\,_{b} \, \varepsilon^{bd}\,
\Big(\,w_{94}\,[A^i\, \Phi^{(1),-}_{\frac{1}{2}\,\,\,\,\,\,\,d}]
+w_{95}\,[G^{-}_{\,\,d} A^i]
\,\Big)\,,
\nonu\\
\Phi^{(1),-}_{\frac{1}{2} \,\,\,\,\,\,\,a} \times \Phi^{(1)}_1
&\sim&
w_{96}\,G^{-}_{\,\,a}+\cdots
+ (\sigma^{i})^{b}\,_a \Big(\,
w_{102}\,[A^i \, \Phi^{(1),-}_{\frac{1}{2}\,\,\,\,\,\,\,b}]
+w_{103}\,[G^{-}_{\,\,b} A^i]\, \Big) \ ,
\nonu\\
\Phi^{(1),-}_{\frac{1}{2} \,\,\,\,\,\,\,a} \times \Phi^{(1),++}_1
&\sim& 
\varepsilon_{ab}\Big(
\,
w_{104}\,G^{+b}
+w_{105}\,\Phi^{(1),+b}_\frac{1}{2}
+w_{106}\,\Phi^{(2),+b}_\frac{1}{2}
+w_{107}\,\Phi^{(1),+b}_\frac{3}{2}
\nonu \\
&  & \quad +\, w_{108}\,[G^{+b} \Phi^{(1)}_0 ] + w_{109}\,[ \Phi^{(1)}_0  \Phi^{(1),+b}_\frac{1}{2}]
\,\Big) 
\nonu\\
& & +\, \varepsilon_{db}\,(\sigma^{i})^{b}\,_{a}
 \Big(\, w_{110}\,[A^i \,\Phi^{(1),+d}_\frac{1}{2} ] +w_{111}\,[G^{+d} A^i] \,\Big) \,  \ ,
\nonu\\
\Phi^{(1),-}_{\frac{1}{2} \,\,\,\,\,\,\, a} \times \Phi^{(1),--}_1
&\sim&
0 \ ,
\nonu\\
\Phi^{(1),+a}_\frac{1}{2}\times \Phi^{(1),i}_1
&\sim&
(\sigma^{i})^{a}\,_{b}\Big(
w_{112}\,G^{+b}
+w_{113}\,\Phi_{\frac{1}{2}}^{(1),+b}
+w_{114}\,\Phi_{\frac{1}{2}}^{(2),+b}
+w_{115}\,\Phi_{\frac{3}{2}}^{(1),+b}
\nonu\\
&
& \qquad +\,  w_{116}\,[\Phi_{0}^{(1)}\Phi_{\frac{1}{2}}^{(1),+b}]
+w_{117}\,[G^{+b}\Phi_{0}^{(1)}]\Big)
\nonu\\
& & +\,  w_{118}\,[G^{+a}A^{i}]
+w_{119}\,[A^{i}\, \Phi_{\frac{1}{2}}^{(1),+a}]
\nonu\\
&
& +\,  \varepsilon_{ijk}(\sigma^{j})^{a}\,_{b}\Big(w_{120}\,[G^{+b}A^{k}]
+w_{121}\,[A^{k}\, \Phi_{\frac{1}{2}}^{(1),+b}]\Big)\, ,
\eea
\bea
\Phi^{(1),-}_{\frac{1}{2}\,\,\,\,\,\,\,a}\times \Phi^{(1),i}_1
&\sim&
(\sigma^{i})^{b}\,_{a} \Big(w_{122}\,G^{-}_{\,\,b}+w_{123}\,\Phi^{(1),-}_{\frac{1}{2}\,\,\,\,\,\,\,b}+w_{124}\,\Phi^{(2),-}_{\frac{1}{2}\,\,\,\,\,\,\,b}
\nonu\\
&
& \qquad +\, w_{125}\,\Phi^{(1),-}_{\frac{3}{2}\,\,\,\,\,\,\,b}+w_{126}\,[\Phi_{0}^{(1)}\Phi^{(1),-}_{\frac{1}{2}\,\,\,\,\,\,\,b}]+w_{127}\,[G^{-}_{\,\,b}\Phi_{0}^{(1)}]\Big)
\nonu\\
&
& +\, w_{128}\,[G^{-}_{\,\,a}A^{i}]+w_{129}\,[A^{i}\, \Phi^{(1),-}_{\frac{1}{2}\,\,\,\,\,\,\,a}]
\nonu\\
&
& +\, (\sigma^{j})^{b}\,_{a}\,\varepsilon_{ijk} \Big(w_{130}\,[G^{-}_{\,\,b}A^{k}]+w_{131}\,[A^{k}\, \Phi^{(1),-}_{\frac{1}{2}\,\,\,\,\,\,\,b}]\Big)\ .
\label{spin7half}
\eea
We have also worked out the most general ansatz for the OPEs of total spin $4$, and the result is given in Appendix~\ref{app:spin4}.

\subsection{Imposing Jacobi identities}

In order to constraint the free parameters we now impose the Jacobi identities associated to the (schematic) OPEs 
%
\bea
&
(G^{+a},\,G^{-}_{\,\,a}) \times \phi^{h_1} \times \phi^{h_2}\ ,\qquad & \hbox{with} \quad 
 h_1+h_2 =  \tfrac{5}{2}\ ,
\nonu\\
&
(L,\,A^i) \times \phi^{h_1} \times \phi^{h_2}\ ,\qquad & \hbox{with} \quad 
 h_1+h_2 = 3\ ,
\nonu\\
& \phi^{h_1} \times \phi^{h_2} \times \phi^{h_3}\ , \qquad & \hbox{with} \quad  
  h_1+h_2+h_3 =  4\ . 
\label{firstJacobiID}
\eea
We find that the only non-zero structure constants in (\ref{Ansatz5H})
and (\ref{Ansatz3}) that are allowed by these Jacobi
identities   are 
\bea
n_ {1} =\frac{c}{3}\,, \quad
w_ {1} = 1\,,\quad
w_ {2} = 1\,, \quad
w_ {17} = 4\,, \quad
w_ {22} = -\frac{2 c}{3}\,, \quad
w_ {24} = -2\,, \quad
w_ {29} = 4\,.
\label{Structualconst1}
\eea
As we mentioned before, the normalisation of $\Phi^{(1)}_0$ associated with $n_1$
in (\ref{Ansatz2}) is not determined by the Jacobi identities; here we have chosen to fix this  
 normalisation constant by setting $w_ {1} = 1$ in (\ref{Ansatz5H}). With this convention, we then have 
$n_1=\frac{c}{3}$ and $w_2=1$.

At this level, the structure constant $w_{11}$ appearing in
(\ref{Ansatz3}) is not yet determined. However, we note that we have the freedom of redefining
the next higher spin generate $ \Phi^{(2)}_{0}$ by 
\bea
\Phi^{(2)}_{0} \mapsto 
\Phi^{(2)}_0+ {\rm const} \,
\bigg[\,[\Phi^{(1)}_0 \Phi^{(1)}_0]-\frac{2(12+c)}{3(9+c)}\,
  L+\frac{4}{(9+c)}\,[A^i A^i]\,\bigg]\ , 
\label{newspin2}
\eea
since also the term in brackets is an ${\cal N}=4$ primary field of spin $s=2$. Using this freedom now allows us to set 
\bea
w_{11}=0 \ , 
\eea
and we shall work with this convention in the following.
\smallskip

In the next step we impose the Jacobi identities arising from 
\bea
&
(G^{+a},\,G^{-}_{\,\,a}) \times \phi^{h_1} \times \phi^{h_2}\ ,\qquad & \hbox{with} \quad 
  h_1+h_2 = 3\,,
\nonu\\
&
(L,\,G^{+a},\,G^{-}_{\,\,a},\,A^i) \times \phi^{h_1} \times \phi^{h_2}\ , \qquad & \hbox{with} \quad  h_1+h_2 = \tfrac{7}{2}\,,
\nonu\\
& \phi^{h_1} \times \phi^{h_2} \times \phi^{h_3}\ , \qquad & \hbox{with} \quad  
 h_1+h_2+h_3 = \tfrac{9}{2}\ .
\label{second jacobi}
\eea
This fixes then the following structure constants 
\bea
w_ {32} & = & \frac {8} {3}\ , \qquad
w_ {34} =1\ , \qquad
w_ {36} = \frac {8 (12 + c)} {(-12 + 8 c +  c^{2})}\ , \qquad
w_ {39} = -\frac {16 c} {(-12 + 8 c + c^{2})}\ , \qquad
\nonu\\
w_ {40} & = & \frac {8} {3}\ , \qquad
w_ {42}= 1\ , \qquad
w_ {44} = \frac {8 (12 + c)} {(-12 + 8 c +  c^{2})}\ , \qquad
w_ {47} = \frac {16 c} {(-12 + 8 c +  c^{2})}\ , 
\eea
\bea
w_ {51} & = & 1\ , \qquad
w_ {53} = -\frac {8 (12 + c)} {(-12 + 8 c + c^{2})}\ , \qquad
w_ {54} = \frac {16 c} {(-12 + 8 c +  c^{2})}\ , \qquad
w_ {59} = 1\ , 
\nonu\\
w_ {61} & = & -\frac {8 (12 +  c)} {(-12 + 8 c + c^{2})}\ , \qquad
w_ {62} = -\frac {16 c} {(-12 + 8 c +  c^{2})}\ , \qquad
w_ {67} = -1\ , 
\nonu\\
w_ {69} & = & \frac {8 (12 + c)} {(-12 + 8 c +  c^{2})} \ , \qquad
w_ {70} = -\frac {16 c} {(-12 + 8 c +  c^{2})}\ , \qquad
w_ {75} = -1\ , 
\nonu\\
w_ {77} & = & \frac {8 (12 + c)} {(-12 + 8 c +   c^{2})}\ , \qquad
w_ {78} = \frac {16 c} {(-12 + 8 c +  c^{2})}\ , \qquad
w_ {80} = -2\ , \qquad
w_ {82} = -\frac{1} {2}\ , 
\nonu\\
w_ {84} & = & -\frac {4 (12 + c)} {(-12 + 8 c + c^{2})}\ , \qquad
w_ {87} = \frac {8 c} {(-12 + 8 c +  c^{2})}\ , \qquad
w_ {88} = 4\ ,  \qquad
w_ {90} = 1\ , 
\nonu\\
w_ {92} & = & \frac {8 (12 + c)} {(-12 + 8 c + c^{2})}\ , \qquad
w_ {95} = -\frac {16 c} {(-12 + 8 c +  c^{2})}\ , \qquad
w_ {96} = 2\ , \qquad
w_ {98} = \frac {1}{2}\ , 
\nonu\\
w_ {100} & = & \frac {4 (12 + c)} {(-12 + 8 c +   c^{2})}\ , \qquad
w_ {103} = \frac {8 c} {(-12 + 8 c +  c^{2})}\ , \qquad
w_ {104} = 4\ , \qquad
w_ {106} = 1\ , 
\nonu\\
w_ {109} & = & \frac {8 (12 + c)} {(-12 + 8 c + c^{2})}\ ,  \qquad
w_ {111} = -\frac {16 c} {(-12 + 8 c +  c^{2})}\ , \qquad
w_ {112} = -1\ , \qquad
w_ {114} = \frac {1} {2}\ , 
\nonu\\
w_ {116} & = & \frac {4 (12 + c)} {(-12 + 8 c + c^{2})}\ , \qquad
w_ {118} = -\frac {8 c} {(-12 + 8 c +  c^{2})}\ , \qquad
w_ {120} = \frac {8\,i\,c} {(-12 + 8 c +  c^{2})}\ , 
\nonu\\
w_ {122} & = & 1\ , \quad
w_ {124} = -\frac {1}{2}\ , \qquad
w_ {126} = -\frac {4 (12 + c)} {(-12 + 8 c +  c^{2})}\ , \qquad
w_ {128} = -\frac {8 c} {(-12 + 8 c + c^{2})}\ , 
\nonu\\
w_ {130} & = & -\frac {8\,i\,c} {(-12 + 8 c + c^{2})}\ .
\label{Structureconst2}
\eea
As before, the normalisation constant $n_2$ of the next lowest higher spin current
$\Phi_{0}^{(2)}$ is related to the structure constant $w_{34}$, and we have used this normalisation freedom to set $w_{34}=1$. Using the Jacobi identity $\Phi^{(2)}_0 \times \Phi^{(2)}_0 \times \Phi^{(2)}_0$, the value of $n_2$ then turns out to be 
\be\label{n2}
n_2 =\frac{16 (-6 + c) c (6 + c)}{9 (-12 + 8 c + c^2)}\ .
\ee
At this stage all the structure constants that appear in (\ref{Ansatz3}) and (\ref{spin7half}) are uniquely determined by consistency; in particular, those that do not appear in (\ref{Structureconst2}) are required to vanish.

In the next step, we have imposed the Jacobi identities corresponding to 
\bea
& (L,\,A^i) \times \Phi^{h_1} \times \Phi^{h_2}\ ,\qquad & \hbox{with} \quad 
 h_1+h_2 = 4\ ,
\nonu\\
& \Phi^{h_1} \times \Phi^{h_2} \times \Phi^{h_3}\ , \qquad & \hbox{with} \quad h_1+h_2+h_3 = 5\ ,
\label{remainJacobi}
\eea
as well as the Jacobi identities arising from 
\bea\label{remainJacobi1}
\de_{ij} \, \Phi^{(1),i}_1 \times \Phi^{(1),j}_1 \times \Phi^{(2)}_0 \ , 
\eea
and
\bea\label{remainJacobi2}
&&\Phi^{(1)}_1 \times \Phi^{(1)}_1 \times \Phi^{(2)}_0\ ,\qquad
\Phi^{(1)}_1 \times \Phi^{(1),i}_1 \times \Phi^{(2)}_0\ ,\qquad
\Phi^{(1),\pm \pm}_1 \times \Phi^{(1),\pm \pm}_1 \times \Phi^{(2)}_0\ ,
\nonu\\
&&\Phi^{(1),\pm \pm}_1 \times \Phi^{(1),i}_1 \times \Phi^{(2)}_0\ ,\quad
\Phi^{(2)}_0 \times \Phi^{(2)}_0 \times \Phi^{(2)}_0\ .
\eea
This then allows us to fix (most) of the structure constants in the ansatz of Appendix~\ref{app:spin4} --- at this stage only the structure constants $w_{182}$ -- $w_{200}$ in the OPE $\Phi_0^{(1)}(z) \,\Phi_0^{(3)}(w)$ of Appendix~\ref{app:structureconstants} are not fixed by this analysis\footnote{One should expect these structure constants to be fixed by the Jacobi identities of total spin $h=\frac{11}{2}$.} --- and the results are spelled out in Appendix~\ref{app:structureconstants}.

\section{Interpretation and consistency checks}\label{sec:4}

The key conclusion from the above analysis is that there are no `low-lying' free parameters that are undetermined by the Jacobi identities except for those one would expect to be fixed by the next set of Jacobi identities. By comparison to what happened say in \cite{Beccaria:2014jra} this therefore suggests that the small ${\cal N}=4$ superconformal ${\cal W}^s_\infty$ algebra with the above spin spectrum does not have any parameter beyond the central charge. This is the main result of our analysis. 
In particular, it suggests that the {\em only} such algebra arises as the $\lambda\rightarrow 0$ limit of the family of large ${\cal N}=4$ ${\cal W}_\infty[\lambda]$ algebras discussed in \cite{Gaberdiel:2014cha}. 


Since this result hinges on the OPE analysis of Section~\ref{sec:3}, it is useful to subject the latter to a number of consistency checks. This is what we shall do in the remainder of this paper.

\subsection{Relation to the large ${\cal N}=4$  ${\cal W}_\infty$ algebra}

The algebra we have analysed from first principles should agree with the $\lambda\rightarrow 0$ limit of the large ${\cal N}=4$  ${\cal W}_\infty[\lambda]$ algebra that was introduced in \cite{Gaberdiel:2014cha}. In terms of the parameters $k_+$ and $k_-$ of the latter algebra, this means that the structure constants of our algebra should arise from those in \cite{Beccaria:2014jra} upon taking $k_+\rightarrow \infty$. In the following we shall confirm that this is indeed the case, once one identifies  correctly how the two constructions are to be identified.

To start with we recall that the linear large ${\cal N}=4$  ${\cal W}_\infty[\lambda]$ algebra of \cite{Gaberdiel:2014cha} contains a free boson and four free fermions that can be decoupled by going to the non-linear description 
\cite{Goddard:1988wv}. These free fields are not present in our small ${\cal N}=4$ ${\cal W}^s_\infty$ algebra, and hence we should compare our results to the non-linear version of the large ${\cal N}=4$ algebra.  The central charge of this non-linear algebra equals 
\be\label{nonlinear}
c_{\rm non-linear} = \frac{ 6 (k_+ +1) (k_- +1)}{k_+ + k_- +2 } - 3 \ , 
\ee 
where $k_\pm$ are the levels of the two $\mathfrak{su}(2)$ algebras in the non-linear algebra.\footnote{Note that the levels of the two $\mathfrak{su}(2)$ algebras in the linear version of the algebra are then $k_\pm + 1$, see the discussion in Section~3.1 of \cite{Gaberdiel:2014cha}. Thus eq.~(\ref{nonlinear}) agrees with eq.~(3.23) of \cite{Gaberdiel:2014cha}.} 

In the limit $k_+\rightarrow \infty$ the currents from the $\mathfrak{su}(2)_{k_+}$ sector of the theory decouple and become three free bosons that are also not present in our small ${\cal N}=4$ ${\cal W}^s_\infty$  description. Thus the central charge of our algebra should agree with 
\be
\lim_{k_+\rightarrow \infty} c_{\rm non-linear} - 3 = 6 k_- \ . 
\ee
Since we identify $k=k_-$, this agrees then with the structure of the small ${\cal N}=4$ superconformal algebra, see in particular the last equation of (\ref{OPELGA}). More explicitly, the relation between the generators that appear in (\ref{OPELGA}) and those of the non-linear large ${\cal N}=4$ algebra of 
\cite{Beccaria:2014jra,AK1509,AKK1703,AKK1910} (denoted by
$T$, $A^{\pm i}$, $G^{\alpha \beta}$ where $\alpha, \beta=1,2$) are 
\bea
& L  = {\displaystyle \lim_{k^+\rightarrow \infty} \Bigl( T+\frac{1}{k_{+}} [A^{+i} \,A^{+i}]  \Bigr)\ ,} \qquad 
& A^i  = i A^{-i}  \ ,  \nonu \\
& {\displaystyle G^{+1}  = \frac{1}{\sqrt{2}}\,G^{22}  \ ,} \qquad 
& G^{+2}  = \frac{1}{\sqrt{2}}\,G^{21}  , \nonu \\
&  {\displaystyle  G^{-}_{\,\,1}  = -\frac{1}{\sqrt{2}}\,G^{11}  \ ,}\qquad 
& G^{-}_{\,\,2}  = \frac{1}{\sqrt{2}}\,G^{12}\ . 
\label{corr}
\eea

For the higher spin currents, the fields $\Phi^{(s)*}_{*}$ of the present paper can be identified with the fields $V^{(s),*}_{*}$ from \cite{Beccaria:2014jra} via
\be\label{dictionary1}
\begin{array}{rclrcl}
 \Phi^{(s)}_0 & = &  i\, g(s)\,V^{(s)}_0 \ , \qquad & &  &
\\[4pt]
{\displaystyle \Phi^{(s),+1}_\frac{1}{2}} & = & {\displaystyle -i\,\frac{g(s)}{\sqrt{2}}\,V^{(s),22}_\frac{1}{2} \ ,} \qquad 
& \Phi^{(s),+2}_\frac{1}{2}  & = & {\displaystyle -i\,\frac{g(s)}{\sqrt{2}}\,V^{(s),21}_\frac{1}{2} \ , }
\\[12pt]
 {\displaystyle \Phi^{(s),-}_{\frac{1}{2}\,\,\,\,\,\,\,1}} &  = & {\displaystyle i\,\frac{g(s)}{\sqrt{2}}\,V^{(s),11}_\frac{1}{2} \ ,} \qquad 
& {\displaystyle \Phi^{(s),-}_{\frac{1}{2}\,\,\,\,\,\,\,2}}   & = & {\displaystyle -i\,\frac{g(s)}{\sqrt{2}}\,V^{(s),12}_\frac{1}{2}  \ ,}
\\[12pt]
 {\displaystyle \Phi^{(s)}_1} &  = & {\displaystyle -\frac{g(s)}{2}\,V^{(s),+3}_1 \ ,}  \qquad & & &
\end{array}
\ee
and
\be\label{dictionary2}
\begin{array}{rclrcl}
 {\displaystyle \Phi^{(s),++}_1} & = & {\displaystyle -\frac{g(s)}{2}\,\big(V^{(s),+1}_1-i\,V^{(s),+2}_1\big) \ , }  \qquad 
 & \Phi^{(s),--}_1 & = & {\displaystyle -\frac{g(s)}{2}\,\big(V^{(s),+1}_1+i\,V^{(s),+2}_1\big) \ , }
\\
 {\displaystyle \Phi^{(s),i}_1} & = & {\displaystyle -\frac{g(s)}{2}\,V^{(s),-i}_1 \ , \quad i=1,2,3} \qquad & &&
\\
 {\displaystyle \Phi^{(s),+1}_\frac{3}{2}} &  = & {\displaystyle i\,\frac{g(s)}{2\sqrt{2}}\,V^{(s),22}_\frac{3}{2}\ , } \qquad 
&  \Phi^{(s),+2}_\frac{3}{2} & = & {\displaystyle i\,\frac{g(s)}{2\sqrt{2}}\,V^{(s),21}_\frac{3}{2} \ , }
\\
{\displaystyle \Phi^{(s),-}_{\frac{3}{2}\,\,\,\,\,\,\,1}} &  = & {\displaystyle -i\,\frac{g(s)}{2\sqrt{2}}\,V^{(s),11}_\frac{3}{2} \ , } 
\qquad 
&  \Phi^{(s),-}_{\frac{3}{2}\,\,\,\,\,\,\,2} & = & {\displaystyle i\,\frac{g(s)}{2\sqrt{2}}\,V^{(s),12}_\frac{3}{2} \ , }
\\
{\displaystyle \Phi^{(s)}_2} &  = & {\displaystyle i\, \frac{g(s)}{4}\,V^{(s)}_2\ , } \qquad   & & &
\end{array}
\ee
where $g(s)$ is a normalisation constant.  If we set $g(s=1)=1$ and $g(s=2)=-\frac{i}{2}$, then the structure constants we have derived above agree precisely with those of \cite{Beccaria:2014jra} in the limit $k^+\rightarrow \infty$. For example, the central term in the OPE 
\bea
\Phi^{(1)}_0 \times \Phi^{(1)}_0 \sim \frac{c}{3}\cdot I\ , 
\label{Spin1Spin1}
\eea
agrees with that from \cite{Beccaria:2014jra}, see eqs.~(3.12) and (3.17) of that paper, since 
\bea
i\,V^{(1)}_0 \times i\,V^{(1)}_0 \Big|_{k_{+} \rightarrow  \infty}
\sim \frac{2\,k_{-}\,k_{+}}{(2+k_{-}+k_{+})}\cdot \, I\,\Big|_{k_{+} \rightarrow  \infty}\sim 2\, k_{-} \cdot I
\ . 
\label{spin1spin1}
\eea
As a slightly more non-trivial example consider the OPE 
\be
\Phi^{(1)}_0  \times  \Phi^{(1),+1}_\frac{1}{2} \sim  G^{+1}\ , 
\ee
where we have used (\ref{Ansatz5H}) with $w_1=1$, see eq.~(\ref{Structualconst1}). Under the above dictionary this matches with (see eqs.~(3.12) and (3.17) of \cite{Beccaria:2014jra})
\be
i\, V^{(1)}_0 \times   \Big(-\frac{i}{\sqrt{2}}\,  V^{(1),22}_\frac{1}{2} \,\Big)
\sim
\frac{1}{\sqrt{2}}\,G^{22} \ , 
\ee
which agrees with our dictionary (\ref{corr}). As an example involving two higher spin currents, we finally consider the last OPE of eq.~(3.15) in \cite{Beccaria:2014jra} 
\bea
V^{(1),\alpha \beta}_\frac{1}{2}(z)\,V^{(1),-i}_1(w)
&=&
\frac{1}{(z-w)^2}\,
c_{1}\,\rho^i_{\gamma\beta}\,G^{\alpha \gamma}(w)
\nonu\\
&& +\, 
\frac{1}{(z-w)}\,\Bigg[\,
\rho^i_{\gamma\beta}\Big(\,
c_{2}\,
V^{(2),\alpha \gamma}_\frac{1}{2}
+c_{3}\,
[V^{(1)}_0  V^{(1),\alpha \gamma}_\frac{1}{2}]
+c_{4}\,
\rho_{j,\,\delta \gamma}\, [A^{-j}G^{\alpha \delta}]
\nonu\\
&& \qquad +\,
c_{5}\,
\rho_{j,\,\delta \alpha}\, [A^{+j}G^{ \delta \gamma}]
+\frac{c_{1}}{3}\,
\partial G^{ \alpha \gamma}
\,\Big)
+c_{7}\, [A^{-i} G^{\alpha \beta}]
\,\Bigg](w)\ ,
\label{BCGexample}
\eea
where the non-vanishing structure constants (that are called $w_{68}$, $w_{71}$, $w_{75}$, $w_{77}$ and $w_{79}$ in \cite{Beccaria:2014jra})\footnote{We are working with the conventions that $\eta^{ij}=\delta^{ij}$, rather than $\eta^{ij} = -\tfrac{1}{2} \delta^{ij}$ as in  \cite{Beccaria:2014jra}; this changes the coefficients of $c_4$ and $c_5$ by a factor $-\tfrac{1}{2}$ relative to  \cite{Beccaria:2014jra}.}
\bea
c_{1} & = & \frac{4(1+2k_{-}+k_{+})}{(2+k_{-}+k_{+})}\ ,\qquad
c_{2}=1\ ,
\\
c_{3}& = &
\frac{4 (k_ {-} - k_ {+}) (5 + 4 k_ {-} + 4 k_ {+} + 2 k_ {-} k_ {+})}{(-4 - 4 k_ {-} - k_ {-}^2 - 4 k_ {+} + 3 k_ {-} k_ {+} + 
  4 k_ {-}^2 k_ {+} - k_ {+}^2 + 4 k_ {-} k_ {+}^2 + 
  3 k_ {-}^2 k_ {+}^2)}\ ,
\nonu\\
c_{4} & = &
\frac{16 (2 + 2 k_ {-} + k_ {+}) (-2 - k_ {-} - k_ {+} + 2 k_ {-} k_ {+} + 
   2 k_ {-} k_ {+}^2)}{(2 + k_ {-} + k_ {+}) (-4 - 4 k_ {-} - k_ {-}^2 - 4 k_ {+} + 
   3 k_ {-} k_ {+} + 4 k_ {-}^2 k_ {+} - k_ {+}^2 + 
   4 k_ {-} k_ {+}^2 + 3 k_ {-}^2 k_ {+}^2)}\ ,
\nonu\\
c_{5} & = &
-\frac{16 (-1 + k_ {-}) (1 + k_ {-}) k_ {+} (2 + 2 k_ {-} + k_ {+})}{(2 + k_ {-} + k_ {+}) (-4 - 4 k_ {-} - k_ {-}^2 - 4 k_ {+} + 
   3 k_ {-} k_ {+} + 4 k_ {-}^2 k_ {+} - k_ {+}^2 + 
   4 k_ {-} k_ {+}^2 + 3 k_ {-}^2 k_ {+}^2)}\ , \nonu
\eea
\be
c_{7}  =  \frac{4}{(2+k_{-}+k_{+})}\ .
\ee
In the limit $k_+\rightarrow \infty$, $c_5 = c_7=0$, and the OPE simplifies to 
\bea
& & V^{(1),\alpha \beta}_\frac{1}{2}(z)\,V^{(1),-i}_1(w) \Big|_{k_{+} \rightarrow \infty} \label{OPEwithk-} \\
&& \qquad = \frac{1}{(z-w)^2}\,
4\,\rho^i_{\gamma\beta}\, G^{\alpha \gamma}(w)
+ \frac{1}{(z-w)}\,\Bigg[\,
\rho^i_{\gamma\beta}\Big(\,
V^{(2),\alpha \gamma}_\frac{1}{2}
-
\frac{8(2+ k_ {-})}{(-1+4 k_{-}+3 k_{-}^2)}\,
[V^{(1)}_0  V^{(1),\alpha \gamma}_\frac{1}{2} ]
\nonu\\
& & \qquad \qquad \qquad +\, 
\frac{32\,k_ {-}}{(-1+4 k_{-}+3 k_{-}^2)}\,
\rho^j_{\delta \gamma}\, [A^{-j}G^{\alpha \delta}]
+ \frac{4}{3}\, 
\partial G^{ \alpha \gamma}
\,\Big)
\,\Bigg](w)\ . \nonumber 
\eea
Translating into our conventions using (\ref{corr}), (\ref{dictionary1}) and (\ref{dictionary2}) and writing $k_- = \frac{c}{6}$, this becomes for $\alpha=\beta=2$ and $i=1$
\bea
&\!\!\!& \Big(-\frac{i}{\sqrt{2}}\,V^{(1),22}_\frac{1}{2} \Big)(z)\,\Big(-\frac{1}{2}\,V^{(1)-1}_1 \Big)(w) \\ 
&\!\!\!& \quad =  -\frac{1}{(z-w)^2}\,
\frac{1}{\sqrt{2}}\,G^{21}(w)
+\,  \frac{1}{(z-w)}\,\Bigg[
-\frac{1}{4\sqrt{2}}\,V^{(2),21}_\frac{1}{2}
+\frac {2 \sqrt{2} (12 + c) } {(-12 + 8 c + c^2)}\,
[V^{(1)}_0 V^{(1),21}_\frac{1}{2}]
\nonu\\
&\!\!\!& \qquad 
-\, \frac{4\,i\, \sqrt{2}\,c}{(-12 + 8 c + c^2)}\,\Big(
[A^{-1}G^{22}] +i\,[A^{-2}G^{22}] -[A^{3}G^{21}]
\Big)
- \frac{1}{3\sqrt{2}}\, 
\partial G^{21}
\Bigg](w)\ , \nonumber 
\eea
which agrees exactly with the OPE
\bea
\Phi^{(1),+1}_\frac{1}{2}(z)\,\Phi^{(1),1}_1(w)\ ,
\nonu
\eea
together with (\ref{spin7half}) and (\ref{Structureconst2}).
We have similarly checked the other OPEs of total spin $\frac{7}{2}$, as well as those of total spin $4$.

\subsection{Truncations}\label{sec:trunc}

Given the relation to the large ${\cal N}=4$ ${\cal W}_\infty[\lambda]$ algebra, we expect that the small ${\cal N}=4$ ${\cal W}^s_\infty$ algebra will truncate for integer values of the central charge,\footnote{There may also be other values of the central charge where the algebra truncates, although given the absence of any triality-like relation \cite{Gaberdiel:2012ku} it is not clear whether other truncations exist.} i.e.\ that the algebra is in fact finitely generated. In terms of the above description, this means that we should expect that some of the higher spin currents can be expressed in terms of normal ordered products of lower generators. Let us parametrise the central charge as $c= 6N$ --- as we shall see this is convenient from the coset perspective --- and let us consider the case $N=1$, corresponding to $c=6$. We then claim that the fields
\be\label{null1}
\Phi^{(1),i}_1 \ , \qquad \Phi^{(1),+a}_\frac{3}{2} \ , \qquad \Phi^{(1),-}_{\frac{3}{2}\,\,\,\,\,\,\,a} \ , \qquad 
\hbox{and} \quad \Phi^{(1)}_2 \ , 
\ee
are in fact equal to normal order products of lower spin fields. The same is also true for all fields from the multiplets $\Phi^{(s)*}_{*}$ with $s\geq 2$.

We can give non-trivial evidence for this claim as follows. We claim that the following combinations are actually null for $N=1$ 
\bea
\varphi^{(1),i}_1&=&\Phi^{(1),i}_1- 2\,[A^i \Phi^{(1)}_0]\ ,
\nonu\\
\varphi^{(1),+a}_\frac{3}{2}&=&\Phi^{(1),+a}_\frac{3}{2}- 2 \,[G^{+a} \Phi^{(1)}_0]+\frac{4}{3}\,(\sigma^i)^{a}\,_{b}\,[A^i \Phi^{(1),+b}_{\frac{1}{2}}]\ ,
\nonu\\
\varphi^{(1),-}_{\frac{3}{2}\,\,\,\,\,\,\,a}&=&\Phi^{(1),-}_{\frac{3}{2}\,\,\,\,\,\,\,a} - 2\,[G^{-}_{\,\,a} \Phi^{(1)}_0]+\frac{4}{3}\,[A^i \Phi^{(1),-}_{\frac{1}{2}\,\,\,\,\,\,\,b}]\,(\sigma^i)^{b}\,_{a}\ ,
\label{nullfields} 
\\
\varphi^{(1)}_2&=&
\Phi^{(1)}_2+ 4\,[L \Phi^{(1)}_0]
+[G^{+a} \Phi^{(1),-}_{\frac{1}{2}\,\,\,\,\,\,\,a}]+[G^{-}_{\,\,a} \Phi^{(1),+a}_{\frac{1}{2}}]
+\frac{4}{3}\,\big[[A^i A^i] \Phi^{(1)}_0\big]+\frac{2}{3}\,
[A^i \Phi^{(1),i}_1] \ .
\nonu
\eea
For example, for the first field we find that the OPE with itself has the form 
\bea
\varphi^{(1),i}_1(z)\,\varphi^{(1),j}_1 (w)
&=&\frac{4 (N-1)N}{(z-w)^4}\,\delta^{ij} +
       {\cal O} \Bigl(\frac{1}{(z-w)^3}\Bigr) \ , 
\eea
whose central term vanishes indeed for $N=1$, thus implying that $\varphi^{(1),i}_1$ is a null field for $N=1$ ($c=6$). Similarly, the central term $n_2$ in the OPE $\Phi^{(2)}_{0}(z)\,\Phi^{(2)}_{0}(w)$, see eq.~(\ref{B1}) in Appendix~\ref{app:spin4}, vanishes at $N=1$ as follows from (\ref{n2}),
\bea\label{Phi2}
\Phi^{(2)}_{0}(z)\,\Phi^{(2)}_{0}(w) & = & \frac{1}{(z-w)^4}\,
  \frac{16c(-36+c^2)}{9(-12+8c+c^2)} +
    {\cal O}\Bigl(\frac{1}{(z-w)^3}\Bigr) \,
    \nonu \\
    & = & \frac{1}{(z-w)^4}\,
      \frac{32(N-1)N(N+1)}{(-1+4N+3N^2)}+
    {\cal O}\Bigl(\frac{1}{(z-w)^3}\Bigr) \, .
\eea
Thus 
$\Phi^{(2)}_{0}$ is null at $N=1$, and the same therefore also applies to the whole multiplet (since the corresponding states can be obtained as ${\cal N}=4$ descendants). 
We have similarly calculated 
\bea
\Phi^{(1),+a}_\frac{1}{2}(z)\,\Phi^{(2)}_0(w)
&\sim &-
\frac{1}{(z-w)}\,\Bigg[\,
  \varphi^{(1),+a}_\frac{3}{2}+\frac{2(N-1)(3N+5)}{(-1+4N+3N^2)}\,
  [G^{+a}\Phi^{(1)}_0]
\nonu\\
&& \qquad -\, 
\frac{4(N-1)(3N+1)}{3(-1+4N+3N^2)}(\sigma^i)^{a}\,_{b}\,
[A^i \Phi^{(1),+b}_\frac{1}{2}]
\,\Bigg](w)\ ,
\nonu\\
\Phi^{(1),-}_{\frac{1}{2}\,\,\,\,\,\,\,a}(z)\,\Phi^{(2)}_0(w)
& \sim &
-\frac{1}{(z-w)}\,\Bigg[\,
\varphi^{(1),-}_{\frac{3}{2}\,\,\,\,\,\,\,a}-\frac{2(N-1)(3N+5)}{(-1+4N+3N^2)}\,[G^{-}_{\,\,\,\,\,\,a}\Phi^{(1)}_0]
\nonu\\
&& \qquad -\,
\frac{4(N-1)(3N+1)}{3(-1+4N+3N^2)}(\sigma^i)^{b}\,_{a}\,
[A^i \Phi^{(1),-}_{\frac{1}{2}\,\,\,\,\,\,\,b}]\,\Bigg](w)\,.
\eea
At $N=1$ only the first term in each case survives, and since $\Phi^{(2)}_0$ is null, it follows that also 
$\varphi^{(1),+a}_\frac{3}{2}$ and $\varphi^{(1),-}_{\frac{3}{2}\,\,\,\,\,\,\,a}$ are null. Finally, in order to see that $\varphi^{(1)}_2$ is null, we compute 
\bea
\Phi^{(1)}_{0}(z)\,\varphi^{(1)}_2 (w) 
&\!\!\!\sim \!\!\!&\frac{1}{(z-w)^2}\,\Bigg[
-4\,\Phi^{(2)}_0
+\frac{4(N-1)(3N+5)}{(-1+4N+3N^2)}\, [\Phi^{(1)}_0 \Phi^{(1)}_0]
\\
&& \!\! +\,
\frac{8(N-1)(-3+2N+3N^2)}{(-1+4N+3N^2)}\,L
+\frac{8(N-1)(1+10 N+3N^2)}{3(-1+4N+3N^2)}\, [A^i A^i]
\,\Bigg](w)\ .
\nonu
\eea
At $N=1$ all terms, except for the first term $(- 4 \Phi^{(2)}_0)$ disappear, but this term is also null because of (\ref{Phi2}). Thus the right-hand-side vanishes, and this therefore implies that also $\varphi^{(1)}_2 $ must be null. As a final consistency check we have evaluated 
\bea
\Phi^{(1),+a}_\frac{1}{2}(z)\,\varphi^{(1),i}_1 (w) 
&=& \frac{1}{(z-w)}\Bigg[
(\sigma^i)^{a}\,_{b}\,\Big\{\,
\frac{1}{2}\,\Phi^{(2),+b}_\frac{1}{2}
{-\frac{(N-1)(3N+5)}{(-1+4N+3N^2)}} [\Phi^{(1)}_0 \Phi^{(1),+b}_\frac{1}{2}] \Big\}
\nonu\\
&& \quad  
+{\frac{2(N+1)(3N-1)}{(-1+4N+3N^2)}}\, [G^{+a}A^i]
\nonu\\
&& \quad +\, 
\frac{4\,i\,N}{(-1+4N+3N^2)}\,\varepsilon_{ijk}\,
(\sigma^j)^{a}\,_{b}\, [G^{+b}A^k]
\,\Bigg](w)\ . 
\eea
At $N=1$ some of the terms disappear manifestly (and $\Phi^{(2),+b}_\frac{1}{2}=0$ because the whole $\Phi^{(2)}$ multiplet vanishes). The remaining combination corresponds to the state 
\begin{eqnarray}
{\cal N}^{ia} & = & \frac{4}{3} \Bigl(  G^{+a}_{-3/2} A^i_{-1} + \frac{i}{2} \epsilon_{ijk} (\sigma^j)^{a}\,_{b}\, G^{+b}_{-3/2} A^k_{-1} \Bigr) |0\rangle \nonumber \\
& = & \frac{4}{3} \Bigl(  A^i_{-1}  G^{+a}_{-3/2} + \frac{i}{2} \epsilon_{ijk} (\sigma^j)^{a}\,_{b}\, A^k_{-1} G^{+b}_{-3/2}  \Bigr) |0\rangle \ , 
\end{eqnarray}
where $|0\rangle$ is the vacuum state. To see that ${\cal N}^{ia}$ is a null state,\footnote{Alternatively, one can also confirm this in the coset approach of Appendix~\ref{app:coset}.} it is enough to show that it is annihilated by all the positive ${\cal N}=4$ superconformal modes.  For example, we find 
\be
\frac{3}{4} A^{l}_{1} \, {\cal N}^{ia} = 
\Big(\frac{c}{12}-\frac{1}{2}\Big)\,\delta^{li}\,G^{+a}_{-3/2} |0\rangle 
+\frac{i}{2}\Big(\frac{c}{12}-\frac{1}{2}\Big)\,\epsilon_{ijl}\,(\sigma^j)^a{}_b\,G^{+b}_{-3/2}|0\rangle \ ,
\ee
which vanishes for $c=6N$ with $N=1$. It is obvious that $G^{+a}_{r} {\cal N}^{ia}  = 0$ for $r>0$, and for the action of $G^{-}_{b,\, 1/2}$ we find 
\be
G^{-}_{b,\, 1/2} \, {\cal N}^{ia} = (\sigma^j)^a{}_b \bigl( 3\, A^i_{-1} A^j_{-1} + 3\, A^j_{-1} A^i_{-1} - 2\,  \delta^{ij} A^l_{-1} A^l_{-1} \bigr) \, |0\rangle \ . 
\ee
One checks that the state on the right-hand-side is indeed null provided that the $\mathfrak{su}(2)_k$ algebra is at level $k=\frac{c}{6}=1$, which is again the case for $c=6$.

\subsection{Truncations in the large ${\cal N}=4$ algebra}

The above truncations do not only occur for the small ${\cal N}=4$ ${\cal W}^s_\infty$ algebra, but actually have a counterpart in the large ${\cal N}=4$ ${\cal W}_\infty[\lambda]$ algebra. This is in fact expected since the case  $c=6N$ with $N$ integer corresponds to the Wolf space coset at integer $N$, which is known to be finitely generated. 

Using the coset terminology (where we parametrise the linear ${\cal N}=4$ ${\cal W}_\infty[\lambda]$ algebra in terms of $N$ and $k$ with $k_+=k+1$ and $k_-=N+1$), the analogue of the null-fields (\ref{nullfields}) at $N=1$ take the form  
%
%
\bea
\varphi^{(1),i}_1&=&\Phi^{(1),i}_1-2\,[A^{i} \Phi^{(1)}_0]
\ , \label{variphispin1} \\
%
\varphi^{(1),+1}_\frac{3}{2}&=&\Phi^{(1),+1}_\frac{3}{2}-2\,[G^{+1} \Phi^{(1)}_0]
+\frac{4(4+k_{+})}{(3+k_{+})}\,[A^{3} \Phi^{(1),+1}_{\frac{1}{2}}]
\nonu\\
&& -\,
\frac{4}{(3+k_{+})}\,\Big([A^{+1} \Phi^{(1),-}_{\frac{1}{2}\,\,\,\,\,\,\,2}]
-i\,[A^{+2} \Phi^{(1),-}_{\frac{1}{2}\,\,\,\,\,\,\,2}]
+[A^{+3} \Phi^{(1)+1}_{\frac{1}{2}}]\Big)\ ,
\label{spin32}\\
\varphi^{(1),+2}_\frac{3}{2}&=&\Phi^{(1),+2}_\frac{3}{2}-2\,[G^{+2} \Phi^{(1)}_0]
-\frac{4(4+k_{+})}{(3+k_{+})}\,[A^{3} \Phi^{(1),+2}_{\frac{1}{2}}]
\nonu\\
&& +\, 
\frac{4}{(3+k_{+})}\,\Big([A^{+1} \Phi^{(1),-}_{\frac{1}{2}\,\,\,\,\,\,\,1}]
-i\,[A^{+2} \Phi^{(1),-}_{\frac{1}{2}\,\,\,\,\,\,\,1}]
-[A^{+3} \Phi^{(1),+2}_{\frac{1}{2}}]\Big)\ ,
\label{spin321}\\
\varphi^{(1),-}_{\frac{3}{2}\,\,\,\,\,\,\,1}&=&\Phi^{(1),-}_{\frac{3}{2}\,\,\,\,\,\,\,1}-2\,[G^{-}_{\,\,1} \Phi^{(1)}_0]
-\frac{4(4+k_{+})}{(3+k_{+})}\,[A^{3} \Phi^{(1),-}_{\frac{1}{2}\,\,\,\,\,\,\,1}]
\nonu\\
&& +\, 
\frac{4}{(3+k_{+})}\,\Big([A^{+1} \Phi^{(1),+2}_{\frac{1}{2}}]
+i\,[A^{+2} \Phi^{(1),+2}_{\frac{1}{2}}]
+[A^{+3} \Phi^{(1),-}_{\frac{1}{2}\,\,\,\,\,\,\,1}]\Big)\ ,
\eea
\bea
\varphi^{(1),-}_{\frac{3}{2}\,\,\,\,\,\,\,2}&=&\Phi^{(1),-}_{\frac{3}{2}\,\,\,\,\,\,\,2}-2\,[G^{-}_{\,\,2} \Phi^{(1)}_0]
+\frac{4(4+k_{+})}{(3+k_{+})}\,[A^{3} \Phi^{(1),-}_{\frac{1}{2}\,\,\,\,\,\,\,2}]
\nonu\\
&& -\,
\frac{4}{(3+k_{+})}\,\Big([A^{+1} \Phi^{(1),+1}_{\frac{1}{2}}]
+i\,[A^{+2} \Phi^{(1),+1}_{\frac{1}{2}}]
-[A^{+3} \Phi^{(1),-}_{\frac{1}{2}\,\,\,\,\,\,\,2}]\Big)\ , 
\label{spin5halfnull} \\
\varphi^{(1)}_2&=&
\Phi^{(1)}_2+4\,[L \Phi^{(1)}_0]
+\frac{2(2+k_{+})}{(3+k_{+})}\,\Big( [G^{+2} \Phi^{(1),-}_{\frac{1}{2}\,\,\,\,\,\,\,2}]+[G^{-}_{\,\,2} \Phi^{(1),+2}_{\frac{1}{2}}]\Big)
-
\frac{2}{(3+k_{+})}\,\Big(
[A_{+1}  \Phi^{(1),++}_1]
\nonu\\
&& + \,
[A_{+1}  \Phi^{(1),--}_1]+i\,[A_{+2}  \Phi^{(1),++}_1]-i\,[A_{+2}  \Phi^{(1),--}_1]
+[A_{+3}  \Phi^{(1)}_1]
+2[A_{+i} A^{+i}  \Phi^{(1)}_0]
\Big)
\nonu\\
&& -\, 
\frac{4(2+k_{+})}{(3+k_{+})}\,[A^{3}\Phi^{(1)}_1]
+\frac{4(5+2k_{+})}{(3+k_{+})}\,[A^{1}A^{1}\Phi^{(1)}_0]
\ . 
\label{nullspin3}
\eea
Let us check that these expressions reduce to the results in (\ref{nullfields}) upon taking 
$k_+\rightarrow\infty$. For example, writing out the Pauli matrices we find from (\ref{nullfields}) 
\bea
\varphi^{(1),+1}_\frac{3}{2} = 
\Phi^{(1),+1}_\frac{3}{2}-2\,[G^{+1} \Phi^{(1)}_0]
+\frac{4}{3}\,\Big([A^1\Phi^{(1),+2}_{\frac{1}{2}}]-i\,[A^2\Phi^{(1),+2}_{\frac{1}{2}}]+[A^3\Phi^{(1),+1}_{\frac{1}{2}}]\Big)\,.
\label{othersmall}
\eea
On the other hand, taking the $k_+\rightarrow \infty$ limit of (\ref{spin32}), we obtain 
\bea
\varphi^{(1),+1}_\frac{3}{2} = \Phi^{(1),+1}_\frac{3}{2}-2\,[G^{+1} \Phi^{(1)}_0]
+4\,[A^3\Phi^{(1),+1}_{\frac{1}{2}}] \ . 
\label{reduced5half}
\eea
The difference between the two expressions 
\bea
  \frac{4}{3} \Bigl( [A^1 \Phi^{(1),+2}_\frac{1}{2}]-i\,
       [A^2 \Phi^{(1),+2}_\frac{1}{2}]-2 \,[A^3 \Phi^{(1),+1}_\frac{1}{2}]
       \Bigr) =0
\label{nullother}
\eea
is actually a null-field at $N=1$, as one can check explicitly in the coset description, see Appendix~\ref{app:coset}. 
A similar analysis can be done for the other null-fields.

In order to see that these fields are null, one method is to work directly in the coset description and confirm that these combinations are indeed zero; this is explained in Appendix~\ref{app:coset}. Alternatively, we can also (with some more effort) show this abstractly. For example, we have calculated the OPE of $\varphi^{(1),+a}_\frac{3}{2}$ with itself 
\bea
\varphi^{(1),+a}_\frac{3}{2}(z)\,
\varphi^{(1),-}_{\frac{3}{2}\,\,\,\,\,\,\,b}(w)
&=&
\frac{\delta^{a}\,_{b}}{(z-w)^5}\,
\frac {32\,(-1 + N) N  k_ {+}} {3 (3 + k_ {+})^2 (2 + N + 
     k_ {+})^3}
\,
 \Big(194 + 190 N + 48 N^2 + 364  k_ {+}
\nonu\\
& &  \qquad + \,
     224 N  k_ {+} + 24 N^2 k_ {+} + 236 k_ {+}^2 + 
     79 N k_ {+}^2 + 3 N^2 k_ {+}^2 \nonu \\
& & \qquad + 63  k_ {+}^3 +  9 N k_ {+}^3 + 6  k_ {+}^4 \Big) + \mathcal{O}\Bigl(\frac{1}{(z-w)^3}\Bigr) \ . 
\eea
In particular, the central term vanishes at $N=1$, thus showing that the field is indeed null. 

In addition to these null fields, also $\Phi^{(2)}_0$ is null at $N=1$ in the large ${\cal N}=4$ algebra. To see this, we can for example compute 
\bea
\Phi^{(1),+a}_\frac{1}{2}(z)\,\Phi^{(2)}_0(w)
&=&
-\frac{1}{(z-w)}\,\Bigg[\,
\varphi^{(1),+a}_\frac{3}{2}+
\\
&& \quad +\,
\frac{4(4+k_{+})}{3(3+k_{+})}(\sigma^i)^{a}\,_{b}\, [A^i \Phi^{(1),+b}_\frac{1}{2}]
-\frac{4(4+k_{+})}{(3+k_{+})} (\sigma^3)^{a}\,_{b}\, [A^3 \Phi^{(1),+b}_\frac{1}{2}]
\,\Bigg](w)\ . 
\nonu
\eea
The term in the first line vanishes, and one can also show that the same is true for the linear combination in the second line; again, this is most easily done using the coset approach, see Appendix~\ref{app:coset}.
\smallskip

A convenient way to obtain the explicit expressions for (\ref{variphispin1}) -- (\ref{nullspin3}) is as follows. (In the following we shall explain the method for $\varphi^{(1),+1}_\frac{3}{2}$, but the same idea can also be applied in the other cases.) For general $N$ and $k$, we can use the explicit form of the OPE relations from \cite{Beccaria:2014jra} to write 
\bea
G^{+1}(z)\, \Phi^{(1)}_1 (w)
&\sim &
-\frac{1}{(z-w)^2}\,\frac{(1+2k_{+}+N)}{(2+k_{+}+N)}\,\Phi^{(1),+1}_\frac{1}{2}(w)
\nonu \\
& & -\, 
\frac{1}{(z-w)}\,\Bigg[\,
\frac{2}{(2+k_{+}+N)}\,\Big(
[A^{1}\Phi^{(1),+2}_{\frac{1}{2}}]-i\,[A^{2}\Phi^{(1),+2}_{\frac{1}{2}}]+[A^{3}\Phi^{(1),+1}_{\frac{1}{2}}]
\nonu \\
& & \quad -\,  [A^{+3}\Phi^{(1),+1}_{\frac{1}{2}}]
\Big)
+\frac{(1+2k_{+}+N)}{3(2+k_{+}+N)}\,\partial  \Phi^{(1),+1}_{\frac{1}{2}}
+\frac{1}{2}\,\Phi^{(1),+1}_{\frac{3}{2}}
\,\Bigg](w)
\ , 
\label{eq1}
\eea
where $A^{+i}$ are the currents associated to $\mathfrak{su}(2)_{k_+}$. On the other hand, we can also evaluate the same OPE for $N=1$ in the coset description of Appendix~\ref{app:coset}, and this leads to 
\bea
G^{+1}(z)\, \Phi^{(1)}_1 (w)
&\sim &
-\frac{1}{(z-w)^2}\,\frac{(2+2k_{+})}{(3+k_{+})}\,\Phi^{(1),+1}_\frac{1}{2}(w)
\nonu \\
& & -\, 
\frac{1}{(z-w)}\,\Bigg[\,
\frac{2}{(3+k_{+})}\,\Big(
[A^{1}\Phi^{(1),+2}_{\frac{1}{2}}]-i\,[A^{2}\Phi^{(1),+2}_{\frac{1}{2}}]+[A^{3}\Phi^{(1),+1}_{\frac{1}{2}}]
\nonu \\
& & \quad -\,  [A^{+3}\Phi^{(1),+1}_{\frac{1}{2}}]
\Big)
+\frac{(2+2k_{+})}{3(3+k_{+})}\,\partial  \Phi^{(1),+1}_{\frac{1}{2}}
\nonu \\
& & \quad +\,  \frac{1}{2}\,
\,\Bigg( 
2\,[G^{+1} \Phi^{(1)}_0]
-\frac{4(4+k_{+})}{(3+k_{+})}\,[A^{3} \Phi^{(1),+1}_{\frac{1}{2}}]
\label{n1opespin5half} \\
&& \quad +\,
\frac{4}{(3+k_{+})}\,\Big([A^{+1} \Phi^{(1),-}_{\frac{1}{2}\,\,\,\,\,\,\,2}]
-i\,[A^{+2} \Phi^{(1),-}_{\frac{1}{2}\,\,\,\,\,\,\,2}]
+[A^{+3} \Phi^{(1)+1}_{\frac{1}{2}}]\Big)
\Bigg) \,\Bigg](w) \ .
\nonu
\eea
Upon comparing the two expressions at $N=1$, we see that the very last term 
in (\ref{eq1}) must agree with the last two lines of (\ref{n1opespin5half}), which is precisely the statement that $\varphi^{(1),+1}_\frac{3}{2}$ as defined in eq.~(\ref{spin32}) is null. The other cases can be dealt with similarly.  

\subsection{The truncation at $N=2$ and beyond}

Returning to the case of the small ${\cal N}=4$ ${\cal W}^s_\infty$ algebra, we have also analysed the corresponding truncation behaviour for $N=2$, using the free field construction (that arises from the coset description in the limit $k_+\rightarrow \infty$). At $N=2$ we have found that the fields 
\be
\Phi^{(2),i}_1 \ , \qquad \Phi^{(2),+a}_\frac{3}{2} \ , \qquad  \Phi^{(2),-}_{\frac{3}{2}\,\,\,\,\,\,\,a}  \ , \qquad \hbox{and} \quad \Phi^{(2)}_{2}
\ee
can be expressed in terms of lower spin fields. (The same is also true for all fields $\Phi^{(s)*}_{*}$ with $s\geq 3$.) This is the natural analogue of the null-relations 
(\ref{variphispin1}) -- (\ref{nullspin3}) for $N=1$. Extrapolating to general $N$, one may therefore be tempted to believe that for integer $N$ the fields 
\be
\Phi^{(N),i}_1 \ , \qquad \Phi^{(N),+a}_\frac{3}{2} \ , \qquad  \Phi^{(N),-}_{\frac{3}{2}\,\,\,\,\,\,\,a}  \ , \qquad \hbox{and} \quad \Phi^{(N)}_{2} \ , 
\ee
as well as all fields $\Phi^{(s)*}_{*}$ with $s\geq N+1$ are not independent, but can in fact be expressed in terms of lower spin fields. 

\section{Conclusions}\label{sec:5}

In this paper we have given strong support for the claim that the symmetry algebra of the symmetric orbifold of $\mathbb{T}^4$ does not have any deformation parameter. More specifically, we have found evidence for the fact that its small ${\cal N}=4$ superconformal ${\cal W}^s_\infty$ subalgebra is uniquely characterised by the central charge, and in particular does not have an additional coupling constant. This suggests in turn that the symmetric orbifold theory has (at least locally) the biggest symmetry algebra in its moduli space. As such it is the natural analogue of free super Yang-Mills in $4$ dimensions. This ties in naturally with the recent observation that it is exactly dual to the tensionless limit of string theory on ${\rm AdS}_3 \times {\rm S}^3 \times \mathbb{T}^4$ \cite{Gaberdiel:2018rqv,Eberhardt:2018ouy,Eberhardt:2019ywk}.

While the symmetric orbifold theory can be analysed using directly the orbifold description, it would also be interesting to get a more direct handle on the theory, using properties of the underlying ${\cal W}^s_\infty$ algebra. For example, it would be interesting to see whether one can find a closed formula for the structure constants of the ${\cal N}=4$ superconformal ${\cal W}^s_\infty$ subalgebra along the lines of \cite{Prochazka:2014gqa}, and whether this can be extended to the full symmetric orbifold algebra. It would also be interesting to find explicit expressions for the eigenvalues of the higher spin currents on the ground states of the single cycle twisted sectors; given the constrained nature of the set-up, one would expect that this should be possible.

\vspace{0.7cm}
\centerline{\bf Acknowledgments}
Man Hea Kim is grateful for a grant from the South Korean-Swiss Young Researchers Exchange Programme (grant NRF-2018K1A3A1A14090664), as well as support from the Pauli Center of ETH Zurich; this enabled him to spend a semester at ETH Zurich. This work of CA was supported by the National Research Foundation of Korea (NRF) grant funded by the Korea government (MSIT) (No.\ 2020R1F1A1066893). CA acknowledges warm hospitality from the School of Liberal Arts (and Institute of Convergence Fundamental Studies), Seoul National University of Science and Technology.
The research of MRG is partially supported by a grant from the Swiss National Science Foundation; his group also acknowledges the support of the NCCR SwissMAP that is funded by the Swiss National Science Foundation.


\appendix

\renewcommand{\theequation}{\Alph{section}\mbox{.}\arabic{equation}}

\section{The OPEs with the supercurrents}\label{app:OPEsuper}

Here we give in detail the OPEs of the spin-$\frac{3}{2}$ supercurrents and the fields from a spin-$s$ multiplet. 
%
\bea
G^{+ a}(z) \, \Phi^{(s)}_0(w) &\sim&-\frac{1}{(z-w)}\, \Phi^{(s),+ a}_\frac{1}{2}(w)\ ,
\nonu\\
G^{-}_{ \,\,a}(z) \, \Phi^{(s)}_0(w) &\sim&-\frac{1}{(z-w)}\, \Phi^{(s),- }_{\frac{1}{2}\,\,\,\,\,\,\, a}(w)\ ,
\nonu\\
G^{+ a}(z) \, \Phi^{(s),+ b}_\frac{1}{2}(w) &\sim&-\frac{1}{(z-w)}\, \varepsilon^{ab}\,\Phi^{(s),++}_1(w) \ ,
\nonu\\
G^{-}_{\,\,a}(z) \, \Phi^{(s),-}_{\frac{1}{2}\,\,\,\,\,\,\,  b}(w) &\sim& -\frac{1}{(z-w)}\, \varepsilon_{ab}\,\Phi^{(s),--}_1(w) \ ,
\nonu\\
G^{+ a}(z) \, \Phi^{(s),- }_{\frac{1}{2}\,\,\,\,\,\,\, b}(w) &\sim&-\frac{1}{(z-w)^2}\, \delta^{a}\,_{b}\,2\,s\,\Phi^{(s)}_0(w)
\nonu \\
&  & +\, 
\frac{1}{(z-w)}\,\Bigg[\delta^{a}\,_{b}  \Big( \Phi^{(s)}_1 - \partial \Phi^{(s)}_0 \Big) + (\sigma^{i})^{a}\,_{b} \, \Phi^{(s),i}_1  \Bigg]   (w)
\ ,
\nonu\\
G^{-}_{ \,\,a}(z) \, \Phi^{(s),+ b}_\frac{1}{2}(w) &\sim&-\frac{1}{(z-w)^2}\, \delta_{a}\,^{b}\,2\,s\,\Phi^{(s)}_0(w)
\nonu \\
& & -\,
\frac{1}{(z-w)}\,\Bigg[\delta_{a}\,^{b}  \Big( \Phi^{(s)}_1 + \partial \Phi^{(s)}_0 \Big) - (\sigma^{i})_{a}\,^{b} \, \Phi^{(s),i}_1  \Bigg]   (w)
\ ,
\nonu\\
G^{+ a}(z) \, \Phi^{(s)}_1(w)&\sim&
- \frac{1}{(z-w)^2}\, (s+1) \,\Phi^{(s),+ a}_\frac{1}{2}(w)
\nonu\\
&& - \, 
 \frac{1}{(z-w)}\, \Big( \frac{1}{2}\, \Phi^{(s),+ a}_\frac{3}{2}+\frac{(s+1)}{(2s+1)}\,\partial \Phi^{(s),+ a}_\frac{1}{2} \Big)(w)\ ,
\nonu\\
G^{- }_{ \,\,a}(z) \, \Phi^{(s)}_1(w)&\sim&
 \frac{1}{(z-w)^2}\, (s+1) \,\Phi^{(s),-}_{\frac{1}{2}\,\,\,\,\,\,\, a}(w)
\nonu\\
&& +\, 
 \frac{1}{(z-w)}\, \Big( \frac{1}{2}\, \Phi^{(s)-}_{\frac{3}{2}\,\,\,\,\,\,\, a}+\frac{(s+1)}{(2s+1)}\,\partial \Phi^{(s),-}_{\frac{1}{2}\,\,\,\,\,\,\, a} \Big)(w)\ ,
\eea
\bea
G^{+ a}(z) \, \Phi^{(s),++ }_1(w)
&\sim& 0\ ,
\nonu\\
G^{-}_{ \,\,a}(z) \, \Phi^{(s),-- }_1(w)
&\sim& 0\ , 
\nonu\\
G^{+ a}(z) \, \Phi^{(s),-- }_1(w)
&\sim&
 \frac{1}{(z-w)^2}\, 2(s+1)\, \varepsilon^{ab}\, \Phi^{(s),-}_{\frac{1}{2}\,\,\,\,\,\,\, b}(w)
\nonu \\
&& +\, 
 \frac{1}{(z-w)}\,\varepsilon^{ab}\, \Big(  \Phi^{(s),-}_{\frac{3}{2}\,\,\,\,\,\,\, b}+\frac{2(s+1)}{(2s+1)}\,\partial \Phi^{(s),-}_{\frac{1}{2}\,\,\,\,\,\,\, b} \Big)(w)\ ,
\nonu\\
G^{-}_{ \,\,a}(z) \, \Phi^{(s),++ }_1(w)
&\sim&
\frac{1}{(z-w)^2}\, 2(s+1)\, \varepsilon_{ab}\, \Phi^{(s),+ b}_\frac{1}{2}(w)
\nonu \\
&& +\, 
 \frac{1}{(z-w)}\,\varepsilon_{ab}\, \Big(  \Phi^{(s),+ b}_\frac{3}{2}+\frac{2(s+1)}{(2s+1)}\,\partial \Phi^{(s),+ b}_\frac{1}{2} \Big)(w)\ ,
\nonu\\
G^{+ a}(z) \, \Phi^{(s),i}_1(w)
&\!\!\sim\!\!&
-\frac{1}{(z-w)^2}\, s\,(\sigma^{i})^{a}\,_{b} \, \Phi^{(s),+ b}_\frac{1}{2}(w)
\nonu \\
&\!\!\!\!& +\, 
\frac{1}{(z-w)}\,(\sigma^{i})^{a}\,_{b} \Big( \frac{1}{2}\, \Phi^{(s),+ b}_\frac{3}{2}-\frac{s}{(2s+1)}\,\partial \Phi^{(s),+ b}_\frac{1}{2} \Big)(w) \ ,
\nonu\\
G^{-}_{ \,\,a}(z) \, \Phi^{(s),i}_1(w)
&\!\!\sim\!\!&
-\frac{1}{(z-w)^2}\,s\,(\sigma^{i})_{a}\,^{b}\, \Phi^{(s),-}_{\frac{1}{2}\,\,\,\,\,\,\, b}(w)  \, 
\nonu \\
&\!\!\!\!& +\, \frac{1}{(z-w)}\,(\sigma^{i})_{a}\,^{b}\Big( \frac{1}{2}\, \Phi^{(s),-}_{\frac{3}{2}\,\,\,\,\,\,\, b}-\frac{s}{(2s+1)}\,\partial \Phi^{(s),-}_{\frac{1}{2}\,\,\,\,\,\,\, b} \Big)(w) \ ,
\nonu\\
G^{+ a}(z) \, \Phi^{(s),+ b }_\frac{3}{2}(w)
&\!\!\sim\!\!&
-\frac{1}{(z-w)^2}\,\frac{4s(s+1)}{(2s+1)}\,\varepsilon^{ab}\,\Phi^{(s),++  }_1(w)
\nonu \\
& \!\!\!\!& -\, 
\frac{1}{(z-w)}\,\frac{2\,s}{(2s+1)}\,\varepsilon^{ab}\,\partial \Phi^{(s),++  }_1 (w)\ ,
\nonu\\
G^{-}_{ \,\,a}(z) \, \Phi^{(s),-}_{\frac{3}{2}\,\,\,\,\,\,\, b}(w)
&\!\!\sim\!\!&
-\frac{1}{(z-w)^2}\,\frac{4s(s+1)}{(2s+1)}\,\varepsilon_{ab}\,\Phi^{(s),-- }_1(w)
\nonu \\
&\!\!\!\! & -\,  \frac{1}{(z-w)}\,\frac{2\,s}{(2s+1)}\,\varepsilon_{ab}\,\partial \Phi^{(s),--  }_1 (w)\ ,
\nonu \\
G^{+ a}(z) \, \Phi^{(s),-}_{\frac{3}{2}\,\,\,\,\,\,\, b}(w)
&\!\!\sim\!\!&
\frac{1}{(z-w)^3}\,\delta^{a}\,_{b}\,\frac{8s(s+1)}{(2s+1)}\, \Phi^{(s)}_0(w)
\nonu\\
&\!\!\!\!& +\, 
\frac{1}{(z-w)^2}\,\Bigg[ \delta^{a}\,_{b}\,\frac{4s(s+1)}{(2s+1)}\, \Phi^{(s)}_1 -\frac{4(1+2s+s^2)}{(2s+1)} \,(\sigma^{i})^{a}\,_{b} \Phi^{(s),i}_1   \Bigg](w)
\nonu\\
&\!\!\!\!& - \, \frac{1}{(z-w)}\,\Bigg[\, \delta^{a}\,_{b}\Big( \Phi^{(s)}_2-\frac{2\,s}{(2s+1)}\, \partial \Phi^{(s)}_1 \Big) 
+\frac{2(s+1)}{(2s+1)} \,(\sigma^{i})^{a}\,_{b}\, \partial \Phi^{(s),i}_1   \,\Bigg] (w)\ ,
\nonu \\
G^{-}_{ \,\,a}(z) \, \Phi^{(s),+ b }_\frac{3}{2}(w)
&\!\!\sim\!\!&
\frac{1}{(z-w)^3}\,\delta_{a}\,^{b}\,\frac{8s(s+1)}{(2s+1)}\, \Phi^{(s)}_0(w)
\nonu\\
&\!\!\!\!& -\,
\frac{1}{(z-w)^2}\,\Bigg[ \delta_{a}\,^{b}\,\frac{4s(s+1)}{(2s+1)}\, \Phi^{(s)}_1 +\frac{4(1+2s+s^2)}{(2s+1)} \,(\sigma^{i})_{a}\,^{b} \Phi^{(s),i}_1   \Bigg](w)
\nonu\\
&\!\!\!\!& -\, \frac{1}{(z-w)}\,\Bigg[\, \delta_{a}\,^{b}\Big( \Phi^{(s)}_2+\frac{2\,s}{(2s+1)}\, \partial \Phi^{(s)}_1 \Big) 
+\frac{2(s+1)}{(2s+1)} \,(\sigma^{i})_{a}\,^{b}\, \partial \Phi^{(s),i}_1   \,\Bigg] (w)\ ,
\nonu
\eea
\bea
G^{+ a}(z) \, \Phi^{(s) }_2(w)
&\!\!\sim\!\!&
-\frac{1}{(z-w)^3}\,\frac{8s(s+1)}{(2s+1)}\, \Phi^{(s),+ a}_\frac{1}{2}(w)
-\frac{1}{(z-w)^2}\,(2s+3)\, \Phi^{(s),+ a}_\frac{3}{2}(w)
\nonu\\
&\!\!\!\!& -\, 
\frac{1}{(z-w)}\, \partial \Phi^{(s),+ a}_\frac{3}{2}(w)\,.
\nonu\\
G^{-}_{ \,\,a}(z) \, \Phi^{(s) }_2(w)
&\!\!\sim\!\!&
-\frac{1}{(z-w)^3}\,\frac{8s(s+1)}{(2s+1)}\,\Phi^{(s),-}_{\frac{1}{2}\,\,\,\,\,\,\,a}(w)
-\frac{1}{(z-w)^2}\,(2s+3)\, \Phi^{(s),-}_{\frac{3}{2}\,\,\,\,\,\,\, a}(w)
\nonu\\
&\!\!\!\!& -\,
\frac{1}{(z-w)}\, \partial \Phi^{(s),-}_{\frac{3}{2}\,\,\,\,\,\,\, a}(w)\ .
\label{OPEGandother}
\eea

\section{Ansatz for total spin $4$}\label{app:spin4}

In this Appendix we present the most general ansatz for the
total spin $4$ case as follows:
\bea\label{B1}
\Phi^{(1)}_0 \times\Phi^{(1)}_2
&\sim&
w_{132}\,\Phi^{(1)}_0
+w_{133}\,L
+w_{134}\,[A^i A^i]
+w_{135}\,[\Phi^{(1)}_0 \Phi^{(1)}_0]
+w_{136}\,\Phi^{(2)}_0 
\nonu \\
& & + \, w_{137}\,\Phi^{(1)}_1 + w_{138}\,\Phi^{(3)}_0
+ w_{139}\,\Phi^{(2)}_1
+w_{140}\,\Phi^{(1)}_2
+w_{141}\,[\Phi^{(1)}_0 \Phi^{(2)}_0]
\nonu\\
& & +\, w_{142}\,[\Phi^{(1)}_0 \Phi^{(1)}_1]
+ w_{143}\,[L \Phi^{(1)}_0]
+w_{144}\,[\Phi^{(1)}_0 \Phi^{(1)}_0 \Phi^{(1)}_0]
\nonu\\
& & +\, w_{145}\,[\Phi^{(1),+a}_\frac{1}{2} \Phi^{(1),-}_{\frac{1}{2}\,\,\,\,\,\,\,a}]
+  w_{146}\,[G^{-}_{\,\,a} \Phi^{(1),+a}_{\frac{1}{2}}]
 + w_{147}\,[G^{+a} \Phi^{(1),-}_{\frac{1}{2}\,\,\,\,\,\,\,a}]
\nonu\\
& & 
+\, w_{148}\,[G^{-}_{\,\,a} G^{+a}]
+w_{149}\,[A^i A^i \Phi^{(1)}_0]
+ w_{150}\,[A^i \Phi^{(1),i}_1]\ ,
\nonu\\
\Phi^{(1)}_0 \times\Phi^{(2)}_1
&\sim&
w_{151}\,\Phi^{(1)}_0 +w_{152}\,L+\cdots
+w_{168}\,[A^i A^i \Phi^{(1)}_0]+w_{169}\,[A^i \Phi^{(1),i}_1]\ ,
\nonu\\
\Phi^{(1)}_0\times \Phi^{(2),++}_1
&\sim&
w_{170}\,\Phi^{(1),++}_1  
+w_{171}\,\Phi^{(2),++}_1  
+w_{172}\,[\Phi^{(1)}_0   \Phi^{(1),++}_1]
\nonu\\
&& +\, \varepsilon_{ab}\,\Big(\,
w_{173}\,[\Phi^{(1),+a}_{\frac{1}{2}} \Phi^{(1),+b}_{\frac{1}{2}} ]
+w_{174}\,[G^{+a} \Phi^{(1),+b}_{\frac{1}{2}} ]
+w_{175}\,[G^{+a} G^{+b} ]
\,\Big)\ ,
\nonu\\
\Phi^{(1)}_0\times \Phi^{(2),--}_1
&\sim &
w_{176}\,\Phi^{(1),--}_1  
+w_{177}\,\Phi^{(2),--}_1  
+w_{178}\,[\Phi^{(1)}_0   \Phi^{(1),--}_1]
\nonu\\
&&
+\, \varepsilon^{ab}\,\Big(\,
w_{179}\,[\Phi^{(1).-}_{\frac{1}{2}\,\,\,\,\,\,\,a} \Phi^{(1).-}_{\frac{1}{2}\,\,\,\,\,\,\,b} ]
+w_{180}\,[G^{-}_{\,\,a} \Phi^{(1).-}_{\frac{1}{2}\,\,\,\,\,\,\,b} ]
+w_{181}\,[G^{-}_{\,\,a} G^{-}_{\,\,b} ]
\,\Big)\ ,
\nonu\\
\Phi^{(1)}_0 \times\Phi^{(3)}_0
&\sim&
w_{182}\,\Phi^{(1)}_0  +w_{183}\,L+\cdots
+w_{199}\,[A^i A^i \Phi^{(1)}_0]+w_{200}\,[A^i \Phi^{(1),i}_1]\ ,
\nonu\\
\Phi^{(1)}_1 \times \Phi^{(1)}_1
&\sim&
w_{201}\,I
+w_{202}\,\Phi^{(1)}_0  +w_{203}\,L+\cdots
+w_{219}\,[A^i A^i \Phi^{(1)}_0]+w_{220}\,[A^i \Phi^{(1),i}_1]\ ,
\nonu\\
\Phi^{(1)}_1 \times \Phi^{(1),++}_1
&\sim&
w_{221}\,\Phi^{(1),++}_1  
+\cdots
+\varepsilon_{ab}\,\Big(\,
\cdots
+w_{226}\,[G^{+a} G^{+b} ]
\,\Big)\ , 
\nonu \\
\Phi^{(1)}_1 \times \Phi^{(1),--}_1
& \sim &
w_{227}\,\Phi^{(1),--}_1  
+\cdots
+\varepsilon^{ab}\,\Big(\,
\cdots
+w_{232}\,[G^{-}_{\,\,a} G^{-}_{\,\,b} ]
\,\Big)\ ,
\nonu\\
\Phi^{(1)}_1 \times \Phi^{(2)}_0
&\sim&
w_{233}\,I
+w_{234}\,\Phi^{(1)}_0  +w_{235}\,L+\cdots
+w_{253}\,[A^i A^i \Phi^{(1)}_0]+w_{254}\,[A^i \Phi^{(1),i}_1]\ ,
\nonu\\
\Phi^{(1),++}_1 \times \Phi^{(1),++}_1
&\sim&
0\ ,
\nonu\\
\Phi^{(1),++}_1 \times \Phi^{(1),--}_1
&\sim&
w_{255}\,I
+w_{256}\,\Phi^{(1)}_0  +w_{257}\,L+\cdots
+w_{273}\,[A^i A^i \Phi^{(1)}_0]+w_{274}\,[A^i \Phi^{(1),i}_1]\ ,
\nonu\\
\Phi^{(1),++}_1 \times \Phi^{(2)}_0
&\sim&
w_{275}\,\Phi^{(1),++}_1  
+\cdots
+\varepsilon_{ab}\,\Big(\,
\cdots
+w_{280}\,[G^{+a} G^{+b} ]
\,\Big)\ ,
\nonu \\
\Phi^{(1),--}_1 \times \Phi^{(1),--}_1
&\sim&
0\  ,
\nonu \\
\Phi^{(1),--}_1 \times \Phi^{(2)}_0
& \sim &
w_{281}\,\Phi^{(1),--}_1  
+\cdots
+\varepsilon^{ab}\,\Big(\,
\cdots
+w_{286}\,[G^{-}_{\,\,a} G^{-}_{\,\,b} ]
\,\Big)\ ,
\eea
\bea
\Phi^{(2)}_0 \times \Phi^{(2)}_0
&\sim&
n_{2}\,I
+w_{287}\,\Phi^{(1)}_0  +w_{288}\,L+\cdots
+w_{304}\,[A^i A^i \Phi^{(1)}_0]+w_{305}\,[A^i \Phi^{(1),i}_1]\ , 
\nonu\\
\Phi^{(1)}_0 \times \Phi^{(2),i}_1
&\sim&
w_{306}\,A^i
+w_{307}\,[A^i \Phi^{(1)}_0]
+w_{308}\,\Phi^{(1),i}_1
+w_{309}\,\Phi^{(2),i}_1
+w_{310}\,[\Phi^{(1)}_0 \Phi^{(1),i}_1]
\nonu\\
&& +\, w_{311}\,[L A^i]
+w_{312}\,[A^i \Phi^{(1)}_0 \Phi^{(1)}_0]
+w_{313}\,[A^i \Phi^{(2)}_0]
+w_{314}\,[A^i \Phi^{(1)}_1]
\nonu\\
&& +\, w_{315}\,[A^i \Phi^{(1)}_0]_{-1}
\nonu \\
& & +\, (\sigma^{i})^{b}\,_{a}
\Big(\,
w_{316}\,[\Phi^{(1),+a}_\frac{1}{2} \Phi^{(1),-}_{\frac{1}{2}\,\,\,\,\,\,\,b}]
+w_{317}\, [G^{+a} \Phi^{(1),-}_{\frac{1}{2}\,\,\,\,\,\,\,b}]
+w_{318}\,[G^{+a} G^{-}_{\,\,b}]
\,\Big) \nonu
\nonu\\
&& +\, w_{319}\,(\sigma^{i})_{b}\,^{a}\,[G^{-}_{\,\,a} \Phi^{(1),+b}_\frac{1}{2}]
+w_{320}\,[A^i A^j A^j]
\nonu\\
& & +\, \varepsilon_{ijk}
\Big(\,w_{321}\,[A^j \Phi^{(1),k}_1]
+w_{322}\,[A^j A^k]_{-1}
\,\Big)\ ,
\nonu\\
\Phi^{(1)}_1 \times \Phi^{(1),i}_1
&\sim&
w_{323}\,A^i
+w_{324}\,[A^i \Phi^{(1)}_0]
+\cdots
+\varepsilon_{ijk}
\Big(\,w_{338}\,[A^j \Phi^{(1),k}_1]
+w_{339}\,[A^j A^k]_{-1}
\,\Big)\ ,
\nonu\\
\Phi^{(1),++}_1 \times \Phi^{(1),i}_1
&\sim&
w_{340}\,[A^i \Phi^{(1),++}_1]
+w_{341}\,\varepsilon_{fb}\,(\sigma^{i})^{b}\,_{a}
\,[G^{+a} \Phi^{(1),+f}_\frac{1}{2}]
\ ,
\nonu \\
\Phi^{(1),--}_1 \times \Phi^{(1),i}_1
& \sim &
w_{342}\,[A^i \Phi^{(1),--}_1]
+w_{343}\,\varepsilon^{fb}\,(\sigma^{i})_{b}\,^{a}
\,[G^{-}_{\,\,a} \Phi^{(1),-}_{\frac{1}{2}\,\,\,\,\,\,\,f}]\ ,
\nonu\\
\Phi^{(1),i}_1 \times \Phi^{(2)}_0
&\sim&
w_{344}\,A^i
+w_{345}\,[A^i \Phi^{(1)}_0]
+\cdots
+\varepsilon_{ijk}
\Big(\,w_{359}\,[A^j \Phi^{(1),k}_1]
+w_{360}\,[A^j A^k]_{-1}
\,\Big)\ ,
\nonu\\
\Phi^{(1),+a}_\frac{1}{2} \times \Phi^{(1),+b}_\frac{3}{2}
&\sim&
\varepsilon^{ab}
\Big\{\,
 w_{361}\,\Phi^{(1),++}_1
 +w_{362}\,\Phi^{(2),++}_1
 +w_{363}\,[\Phi^{(1)}_0 \Phi^{(1),++}_1]
\nonu\\
& & +\, \varepsilon_{ed}\Big(\,
w_{364}\,[\Phi^{(1),+d}_{\frac{1}{2}} \Phi^{(1),+e}_\frac{1}{2}]
 +w_{365}\,[G^{+d} \Phi^{(1),+e}_\frac{1}{2}]
+w_{366}\,[G^{+d} G^{+e} ]
\,\Big) \Big\}
\nonu\\
&& +\, (\sigma^{i})^{ab}\,\Big\{\,w_{367}\,[A^i \Phi^{(1),++}_1]
+\varepsilon_{fd}\,(\sigma^{i})^{d}\,_{e}\, w_{368}\, [G^{+e} \Phi^{(1),+f}_\frac{1}{2}]
\Big\}\ ,
\nonu\\
\Phi^{(1),+a}_\frac{1}{2} \times \Phi^{(1),-}_{\frac{3}{2}\,\,\,\,\,\,\,b}
&\sim&
\delta^{a}\,_{b}
\Big\{\,
w_{369}\,\Phi^{(1)}_0 +\cdots
+w_{387}\,[A^i \Phi^{(1),i}_1]
\,\Big\}
\nonu\\
&& +\, (\sigma^{i})^{a}\,_{b}
\Big\{\,
w_{388}\,A^i+
\cdots
+\varepsilon_{ijk} \Big(\,
w_{403}\,[A^j \Phi^{(1),k}_1]
+w_{404}\,[A^j  A^k]_{-1}
\,\Big)
\Big\}\ ,
\nonu\\
\Phi^{(1),-}_{\frac{1}{2}\,\,\,\,\,\,\,a} \times \Phi^{(1),+b}_\frac{3}{2}
&\sim&
\delta_{a}\,^{b}
\Big\{\,
w_{405}\,\Phi^{(1)}_0 +\cdots
+w_{423}\,[A^i \Phi^{(1),i}_1]
\,\Big\}
\nonu\\
&& +\,
(\sigma^{i})_{a}\,^{b}
\Big\{\,
w_{424}\,A^i+
\cdots
+\varepsilon_{ijk} \Big(\,
w_{439}\,[A^j \Phi^{(1),k}_1]
+w_{440}\,[A^j  A^k]_{-1}
\,\Big)
\Big\}\ ,
\nonu\\
\Phi^{(1),-}_{\frac{1}{2}\,\,\,\,\,\,\,a} \times \Phi^{(1),-}_{\frac{3}{2}\,\,\,\,\,\,\,b}
&\sim&
\varepsilon_{ab}
\Big\{\,
w_{441}\,\Phi^{(1),--}_1
 +w_{442}\,\Phi^{(2),--}_1
  +w_{443}\,[\Phi^{(1)}_0 \Phi^{(1),--}_1]
 \nonu\\
&&
+\varepsilon^{ed}\,\Big(\,
w_{444}\,[\Phi^{(1),-}_{\frac{1}{2}\,\,\,\,\,\,\,d} \Phi^{(1),-}_{\frac{1}{2}\,\,\,\,\,\,\,e}]
 +w_{445}\,[G^{-}_{\,\,d} \Phi^{(1),-}_{\frac{1}{2}\,\,\,\,\,\,\,e}]
+w_{446}\,[G^{-}_{\,\,d} G^{-}_{\,\,e}]
\,\Big)
\, \Big\}
 \nonu\\
&&
+\, (\sigma^{i})_{ab}\,
\Big\{\,
w_{447}\,[A^i \Phi^{(1),--}_1]
+\varepsilon^{fd}\,(\sigma^{i})_{d}\,^{e}\, w_{448}\, [G^{-}_{\,\,e} \Phi^{(1),-}_{\frac{1}{2}\,\,\,\,\,\,\,f}]
\, \Big\}
\ ,
\nonu \\
\Phi^{(1),+a}_\frac{1}{2} \times \Phi^{(2),+b}_\frac{1}{2}
&\sim &
\varepsilon^{ab}
\Big\{\,
w_{449}\,\Phi^{(1),++}_1 +\cdots
\,\Big\}
+
(\sigma^{i})^{ab}\,
\Big\{\,
\cdots
+\varepsilon_{fd}\,(\sigma^{i})^{d}\,_{e}\, w_{456}\, [G^{+e} \Phi^{(1),+f}_\frac{1}{2}]
\, \Big\} \, ,
\nonu\\
\Phi^{(1),+a}_\frac{1}{2} \times \Phi^{(2),-}_{\frac{1}{2}\,\,\,\,\,\,\,b}
&\sim&
\delta^{a}\,_{b}
\Big\{\,
w_{457}\,\Phi^{(1)}_0 +\cdots
+w_{475}\,[A^i \Phi^{(1),i}_1]
\,\Big\}
\nonu\\
&
+&
(\sigma^{i})^{a}\,_{b}
\Big\{\,
w_{476}\,A^i+
\cdots
+\varepsilon_{ijk} \Big(\,
w_{491}\,[A^j \Phi^{(1),k}_1]
+w_{492}\,[A^j  A^k]_{-1}
\,\Big)
\Big\}\ ,
\nonu\\
\Phi^{(1),-}_{\frac{1}{2}\,\,\,\,\,\,\,a} \times \Phi^{(2),+b}_\frac{1}{2}
&\sim&
\delta^{b}\,_{a}
\Big\{\,
w_{493}\,\Phi^{(1)}_0 +\cdots
+w_{511}\,[A^i \Phi^{(1),i}_1]
\,\Big\}
\nonu\\
&& +\,
(\sigma^{i})_{a}\,^{b}
\Big\{\,
w_{512}\,A^i+
\cdots
+\varepsilon_{ijk} \Big(\,
w_{527}\,[A^j \Phi^{(1),k}_1]
+w_{528}\,[A^j  A^k]_{-1}
\,\Big)
\Big\}\ ,
\eea
\bea
\Phi^{(1),-}_{\frac{1}{2}\,\,\,\,\,\,\,a} \times \Phi^{(2),-}_{\frac{1}{2}\,\,\,\,\,\,\,b}
&\sim&
\varepsilon_{ab}
\Big\{\,
w_{529}\,\Phi^{(1),--}_1 +\cdots
\,\Big\}
+
(\sigma^{i})_{ab}\,
\Big\{\,
\cdots
+\varepsilon^{fd}\,(\sigma^{i})_{d}\,^{e}\, w_{538}\, [G^{-}_{\,\,e} \Phi^{(1),-}_{\frac{1}{2}\,\,\,\,\,\,\,f}]
\, \Big\}\ , \nonu \\
\Phi^{(1),i}_1\times \Phi^{(1),j}_1
&\sim&
\eta^{ij}
\Big\{\,
w_{539}\,I
+w_{540}\,L
+w_{541}\,[A^k A^k]
+w_{542}\,[\Phi^{(1)}_0 \Phi^{(1)}_0]
+w_{543}\,\Phi^{(2)}_0
\nonu\\
&& \quad +\,  w_{544}\,\Phi^{(1)}_1
+w_{545}\,\Phi^{(3)}_0
+w_{546}\,\Phi^{(2)}_1
\nonu\\
&& \quad +\, w_{547}\,\Phi^{(1)}_2
+w_{548}\,[\Phi^{(1)}_0 \Phi^{(2)}_0]
+w_{549}\,[\Phi^{(1)}_0 \Phi^{(1)}_1]
\nonu\\
&& \quad +\, w_{550}\,[L \Phi^{(1)}_0 ]
+w_{551}\,[\Phi^{(1)}_0 \Phi^{(1)}_0  \Phi^{(1)}_0 ]
\nonu\\
&& \quad +\, w_{552}\,[\Phi^{(1),+a}_{\frac{1}{2}} \Phi^{(1),-}_{\frac{1}{2}\,\,\,\,\,\,\,a}]
+w_{553}\,[G^-_{\,\,a} \Phi^{(1),+a}_{\frac{1}{2}}]
+w_{554}\,[G^{+a} \Phi^{(1),-}_{\frac{1}{2}\,\,\,\,\,\,\,a}]
\nonu\\
&& \quad +\, w_{555}\,[G^{-}_{\,\, a} G^{+a}]
+w_{556}\,[A^k A^k \Phi^{(1)}_0]
+w_{557}\,[A^k \Phi^{(1),k}_1]\,\Big\}
\nonu\\
&& +\, \varepsilon_{ijk}\,\Big\{\,
w_{558}\,A^k
+w_{559}\,[A^k \Phi^{(1)}_0]
+w_{560}\,\Phi^{(1),k}_1
+w_{561}\, \Phi^{(2),k}_1
\nonu \\
& & \quad  +\,  w_{562}\,[ \Phi^{(1)}_0  \Phi^{(1),k}_1 ]
+ w_{563}\, [L A^k]
+w_{564}\,[A^k  \Phi^{(1)}_0  \Phi^{(1)}_0]
+w_{565}\,[A^k  \Phi^{(2)}_0] 
\nonu\\
&& \quad +\, w_{566}\,[A^k  \Phi^{(1)}_1]
+ w_{567}\,[A^k  \Phi^{(1)}_0]_{-1}
\nonu\\
&& \quad 
+\, (\sigma^{k})^{b}\,_{a}
\Big(\,
w_{568}\,[\Phi^{(1),+a}_\frac{1}{2} \Phi^{(1),-}_{\frac{1}{2}\,\,\,\,\,\,\,b}]
+w_{569}\, [G^{+a} \Phi^{(1),-}_{\frac{1}{2}\,\,\,\,\,\,\,b}]
+w_{570}\,[G^{+a} G^{-}_{\,\,b}]
\,\Big)  
\nonu\\
&& \quad +\, w_{571}\,[G^{-}_{\,\,a} \Phi^{(1),+b}_\frac{1}{2}]\,(\sigma^{k})_{b}\,^{a} + w_{572}\,[A^k A^l A^l]
\nonu\\
&& \quad +\, \varepsilon_{klm}
\Big(\,w_{573}\,[A^l \Phi^{(1),m}_1] +\, w_{574}\,[A^l A^m]_{-1}
\,\Big)\Big\}
\nonu\\
&& 
+w_{575}\,[A^i A^j]
+w_{576}\,[A^i  \Phi^{(1),j}_1]
+w_{577}\,[A^j  \Phi^{(1),i}_1] +\,  w_{578}\,[A^i  A^j]_{-1} \ . 
\eea

\section{Structure constants of the OPEs}\label{app:structureconstants}

The various structure constants appearing in Appendix~\ref{app:spin4} are determined by the Jacobi identities (\ref{remainJacobi}), (\ref{remainJacobi1}) and (\ref{remainJacobi2}), and are explicitly given as 
\bea
&&w_ {133} = -\frac {16 (-6 + c) (6 + c)} {3 (-12 + 8 c + c^{2})}\ ,
\quad  
w_ {134} = -\frac {64 c} {(-12 + 8 c + c^{2})}\ , \quad  
w_ {135} = -\frac {16 (12 + c)} {(-12 + 8 c + c^{2})}\ ,
\nonu\\
&&
w_ {136} = -4\ , \quad  
w_ {156} =- \frac {8 (-6 + c) (6 + c)} {3 (-12 + 8 c + c^{2})}\ , \quad  
w_ {165} = \frac {8 (-6 + c)} {(-12 + 8 c + c^{2})}\ , \quad  
\nonu\\
&&
w_ {166} = -\frac {8 (-6 + c)} {(-12 + 8 c + c^{2})}\ , \quad  
w_ {170} = -\frac {8 (-6 + c) (6 + c)} {3 (-12 + 8 c +  c^{2})}\ , \quad  
w_ {174} = \frac {16 (-6 + c)} {(-12 + 8 c + c^{2})}\ , \quad  
\nonu\\
&&
w_ {176} = -\frac {8 (-6 + c) (6 + c)} {3 (-12 + 8 c + c^{2})}\ , \quad  
w_ {180} = \frac {16 (-6 + c)} {(-12 + 8 c + c^{2})}\ , \quad  
w_ {201} = \frac {4 c} {3}\ , \quad  
\nonu\\
&&
w_ {203} = \frac {16 c (12 + c)} {3 (-12 + 8 c + c^{2})}\ , \quad  
w_ {204} = -\frac {32 c} {(-12 + 8 c + c^{2})}\ , \quad  
w_ {205} = -\frac {8 (12 + c)} {(-12 + 8 c + c^{2})}\ , \quad  
\nonu\\
&&
w_ {206} = -2\ , \quad  
w_ {207} = 1\ , \quad  
w_ {223} = \frac {8 (12 + c)} {(-12 + 8 c + c^{2})}\ , \quad  
w_ {224} = \frac {4 (12 + c)} {(-12 + 8 c + c^{2})}\ , \quad  
\eea
\bea
&&
w_ {226} = -\frac {12 c} {(-12 + 8 c + c^{2})}\ , \quad  
w_ {228} = -1\ , \quad  
w_ {229} = -\frac {8 (12 + c)} {(-12 + 8 c + c^{2})}\ , \quad  
\nonu\\
&&
w_ {230} = -\frac {4 (12 + c)} {(-12 + 8 c + c^{2})}\ , \quad  
w_ {232} = \frac {12 c} {(-12 + 8 c + c^{2})}\ , \quad  
w_ {239} = -\frac {8 (-6 + c) (6 + c)} {3 (-12 + 8 c + c^{2})}\ , \quad  
\nonu\\
&&
w_ {255} = \frac {8 c} {3}\ , \quad  
w_ {257} = \frac {32 c (12 + c)} {3 (-12 + 8 c + c^{2})}\ , \quad  
w_ {258} = -\frac {64 c} {(-12 + 8 c + c^{2})}\ , \quad  
\nonu\\
&&
w_ {259} = -\frac {16 (12 + c)} {(-12 + 8 c + c^{2})}\ , \quad  
w_ {260} = -4\ , \quad  
w_ {263} = 2\ , \quad  
w_ {266} = \frac {16 (12 + c)} {(-12 + 8 c + c^{2})}\ , \quad  
\nonu\\
&&
w_ {269} = -\frac {8 (12 + c)} {(-12 + 8 c + c^{2})}\ , \ \   
w_ {272} = -\frac {24 c} {(-12 + 8 c + c^{2})}\ , \ \   
w_ {275} = -\frac {8 (-6 + c) (6 + c)} {3 (-12 + 8 c + c^{2})}\ , 
\nonu\\
&&
w_ {281} = -\frac {8 (-6 + c) (6 + c)} {3 (-12 + 8 c + c^{2})}\ , \quad  
w_ {288} = \frac {64 c (12 + c) (-6 + c) (6 + c)} {9 (-12 + 8 c + c^{2})^{2}}\ , \quad  
\nonu\\
&&
w_ {289} = -\frac {128 c (-6 + c) (6 + c)} {3 (-12 + 8 c + c^{2})^{2}}\ , \quad 
w_ {290} = -\frac {32 (12 + c) (-6 + c) (6 + c)} {3 (-12 + 8 c + c^{2})^{2}}\ , \quad 
\nonu\\
&&
w_ {291} = \frac {8 (36 + 24 c + c^{2})} {3 (-12 + 8 c + c^{2})}\ ,\quad 
w_ {307} = -\frac {64 (6 + c)} {(-12 + 8 c + c^{2})}\ , \quad  
w_ {308} = \frac {16 c (6 + c)} {3 (-12 + 8 c + c^{2})}\ , \quad  
\nonu\\
&&
w_ {317} = \frac {8 (6 + c)} {(-12 + 8 c + c^{2})}\ , \quad  
w_ {319} = \frac {8 (6 + c)} {(-12 + 8 c + c^{2})}\ , \quad  
w_ {345} = -\frac {64 (6 + c)} {(-12 + 8 c + c^{2})}\ , \quad  
\nonu\\
&&
w_ {346} = \frac {16 c (6 + c)} {3 (-12 + 8 c + c^{2})}\ , \quad  
w_ {359} = -\frac {32\, i\, c} {(-12 + 8 c + c^{2})}\ , \quad  
w_ {362} = 1\ , \quad  
\nonu\\
&&
w_ {363} = \frac {8 (12 + c)} {(-12 + 8 c + c^{2})}\ , \quad  
w_ {364} = \frac {4 (12 + c)} {(-12 + 8 c + c^{2})}\ , \quad  
w_ {366} = -\frac {12 c} {(-12 + 8 c + c^{2})}\ , \quad  
\nonu\\
&&
w_ {370} = -\frac {128 (3 + c)} {3 (-12 + 8 c + c^{2})}\ , \quad  
w_ {371} = \frac {64 c} {(-12 + 8 c + c^{2})}\ , \quad  
w_ {372} = \frac {16 (12 + c)} {(-12 + 8 c + c^{2})}\ , \quad  
\nonu\\
&&
w_ {373} = 4\ , \quad  
w_ {376} = -1\ , \quad  
w_ {379} = -\frac {8 (12 + c)} {(-12 + 8 c + c^{2})}\ , \quad  
w_ {382} = \frac {4 (12 + c)} {(-12 + 8 c + c^{2})}\ , \quad  
\nonu\\
&&
w_ {385} = \frac {12 c} {(-12 + 8 c + c^{2})}\ , \quad  
w_ {388} = -\frac {32} {3}\ , \quad  
w_ {391} = -1\ , \quad  
w_ {392} = -\frac {8 (12 + c)} {(-12 + 8 c + c^{2})}\ , \quad  
\nonu\\
&&
w_ {393} = -\frac {32 c} {(-12 + 8 c + c^{2})}\ , \quad  
w_ {398} = \frac {4 (12 + c)} {(-12 + 8 c + c^{2})}\ , \quad  
w_ {400} = \frac {4 c} {(-12 + 8 c + c^{2})}\ , \quad  
\nonu\\
&&
w_ {406} = -\frac {128 (3 + c)} {3 (-12 + 8 c + c^{2})}\ , \quad  
w_ {407} = \frac {64 c} {(-12 + 8 c + c^{2})}\ , \quad  
w_ {408} = \frac {16 (12 + c)} {(-12 + 8 c + c^{2})}\ , \quad  
\nonu\\
&&
w_ {409} = 4\ , \quad  
w_ {412} = 1\ , \quad  
w_ {415} = \frac {8 (12 + c)} {(-12 + 8 c + c^{2})}\ , \quad  
w_ {418} = -\frac {4 (12 + c)} {(-12 + 8 c + c^{2})}\ , \quad  
\nonu\\
&&
w_ {421} = -\frac {12 c} {(-12 + 8 c + c^{2})}\ , \quad  
w_ {424} = -\frac {32} {3}\ , \quad  
w_ {427} = -1\ , \quad  
w_ {428} = -\frac {8 (12 + c)} {(-12 + 8 c + c^{2})}\ , 
\nonu\\
&&
w_ {429} = -\frac {32 c} {(-12 + 8 c + c^{2})}\ , \quad  
w_ {430} = \frac {4 (12 + c)} {(-12 + 8 c + c^{2})}\ , \quad  
w_ {436} = \frac {(4 c)} {(-12 + 8 c + c^{2})}\ , \quad  
\eea
\bea
&&
w_ {442} = 1, \quad  w_ {443} = \frac {8 (12 + c)} {(-12 + 8 c + c^{2})}\ , \quad  
w_ {444} = \frac {4 (12 + c)} {(-12 + 8 c + c^{2})}\ , \quad  
\nonu\\
&&
w_ {446} = -\frac {12 c} {(-12 + 8 c + c^{2})}\ , \quad  
w_ {449} = \frac {8 (-6 + c) (6 + c)} {3 (-12 + 8 c + c^{2})}\ , \quad  
w_ {453} = -\frac {8 (-6 + c)} {(-12 + 8 c + c^{2})}\ , \quad  
\nonu\\
&&
w_ {455} = -\frac {16 c} {(-12 + 8 c + c^{2})}\ , \quad
w_ {456} = \frac {48} {(-12 + 8 c + c^{2})}\ , \quad  
w_ {462} = -\frac {8 (-6 + c) (6 + c)} {3 (-12 + 8 c + c^{2})}\ , \quad  
\nonu \\
&&
w_ {465} = 1\ , \quad  
w_ {468} = \frac {16 (12 + c)} {(-12 + 8 c + c^{2})}\ , \quad  
w_ {471} = \frac {12 c} {(-12 + 8 c + c^{2})}\ , \quad  
\nonu \\
&&
w_ {472} = \frac {4 (12 + c)} {(-12 + 8 c + c^{2})}\ , \quad  
w_ {475} = \frac {16 c} {(-12 + 8 c + c^{2})}\ , \quad  
w_ {477} = -\frac {64 (6 + c)} {(-12 + 8 c + c^{2})}\ , \quad  
\nonu \\
&&
w_ {478} = \frac {16 c (6 + c)} {3 (-12 + 8 c + c^{2})}\ , \quad  
w_ {484} = \frac {16 c} {(-12 + 8 c + c^{2})}\ , \quad  
w_ {487} = \frac {4 (12 + c)} {(-12 + 8 c + c^{2})}\ , \quad  
\nonu\\
&&
w_ {489} = \frac {4 c} {(-12 + 8 c + c^{2})}\ , \quad  
w_ {491} = -\frac {16\, i\, c} {(-12 + 8 c + c^{2})}\ , \quad  
w_ {498} = \frac {8 (-6 + c) (6 + c)} {3 (-12 + 8 c + c^{2})}\ , \quad  
\nonu\\
&&
w_ {501} = 1\ , \quad  
w_ {504} = \frac {16 (12 + c)} {(-12 + 8 c + c^{2})}\ , \quad  
w_ {507} = \frac {4 (12 + c)} {(-12 + 8 c + c^{2})}\ , \quad  
\nonu\\
&&
w_ {508} = \frac {12 c} {(-12 + 8 c + c^{2})}\ , \quad  
w_ {511} = \frac {16 c} {(-12 + 8 c + c^{2})}\ , \quad  
w_ {513} = -\frac {64 (6 + c)} {(-12 + 8 c + c^{2})}\ , \quad  
\nonu\\
&&
w_ {514} = \frac {16 c (6 + c)} {3 (-12 + 8 c + c^{2})}\ , \quad  
w_ {520} = -\frac {16 c} {(-12 + 8 c + c^{2})}\ , \quad  
w_ {523} = \frac {4 c} {(-12 + 8 c + c^{2})}\ , \quad  
\nonu\\
&&
w_ {525} = \frac {4 (12 + c)} {(-12 + 8 c + c^{2})}\ , \quad  
w_ {527} = -\frac {16\, i\, c} {(-12 + 8 c + c^{2})}\ , \quad  
w_ {529} = -\frac {8 (-6 + c) (6 + c)} {3 (-12 + 8 c +  c^{2})}\ , \quad  
\nonu\\
&&
w_ {533} = \frac {8 (-6 + c)} {(-12 + 8 c + c^{2})}\ , \quad  
w_ {537} = -\frac {16 c} {(-12 + 8 c + c^{2})}\ , \quad  
w_ {538} = \frac {48} {(-12 + 8 c + c^{2})}\ , \quad  
\\
&&
w_ {540} = \frac {8 (-6 + c) (6 + c)} {3 (-12 + 8 c + c^{2})}\ , \quad  
w_ {541} = \frac {32 c} {(-12 + 8 c + c^{2})}\ , \quad  
w_ {542} = \frac {8 (12 + c)} {(-12 + 8 c + c^{2})}\ , \quad  
\nonu \\
&&
w_ {543} = 2\ , \quad  
w_ {558} = 8 i\ , \quad  
w_ {561} = i\ , \quad  
w_ {562} = \frac {8 i (12 + c)} {(-12 + 8 c + c^{2})}\ , \quad  
\nonu\\
&&
w_ {563} = \frac {32\, i\, c} {(-12 + 8 c + c^{2})}\ , \quad  
w_ {568} = -\frac {4 i (12 + c)} {(-12 + 8 c + c^{2})}\ , \quad  
w_ {570} = -\frac {4\, i\, c} {(-12 + 8 c + c^{2})}\ . 
\nonu
\eea

\section{The coset description at $N=1$}\label{app:coset}

Some of the calculations of Section~\ref{sec:trunc} are most easily done in the coset description. Using the conventions of \cite{AK1509}, where $V^a(z)$ are the superaffine currents of $\mathfrak{su}(N+2)_k$, while $Q^a(z)$ are the corresponding fermions in the adjoint representation of $\mathfrak{su}(N+2)$, we can write the fields of the Wolf space coset theory at $N=1$ as 
\bea
L(z)
&=&
\frac{3}{2(3+k_{+})^3}\,Q^1 Q^2 Q^5 Q^6(z) + \frac{1}{(3+k_{+})}\,(V^1 V^5+V^2 V^6)(z)
\label{Lexp} \\
&& -\, 
\frac{1}{4(3+k_{+})^3}\,
\Big(
(3i+\sqrt{3})(Q^1 Q^5 V^4+Q^2 Q^6 V^4)) \nonu \\ 
& & \qquad  \qquad
+(-3i+\sqrt{3})(Q^1 Q^5 V^8+Q^2 Q^6 V^8))
\nonu\\
&& \qquad \qquad +\, 
(3+2k_{+})(\partial Q^1 Q^5+\partial Q^2 Q^6-Q^1 \partial Q^5-Q^2 \partial Q^6 )
\Big)(z) 
\nonu\\
&& -\, 
\frac{1}{2k_{+}(3+k_{+})}\,\Big(
6\,V^3 V^7+3\,V^4 V^8
+\frac{3(1-i\sqrt{3})}{4}\,V^4 V^4
+\frac{3(1+i\sqrt{3})}{4}\,V^8 V^8
\nonu \\
&& -\, 
\frac{1}{2}
(3i-3\sqrt{3}+3i\,k_{+}+\sqrt{3} k_{+})\,\partial V^4
-\frac{1}{2}
(-3i-3\sqrt{3}-3i\,k_{+}+\sqrt{3} k_{+})\,\partial V^8
\Big)(z)\ .
\nonu
\eea
The four spin-$\frac{3}{2}$ currents are given by
\bea
G^{+1}(z)
&\!\!=\!\!&
-\frac{\sqrt{2}}{(3+k_{+})}\,\Big(
Q^1 V^5+Q^6 V^2
\Big)(z)\ , \ \ 
G^{+2}(z)
=
\frac{\sqrt{2}}{(3+k_{+})}\,\Big(
Q^2 V^5-Q^5 V^2
\Big)(z)\ ,
\nonu\\
G^{-}_{\,\,1}(z)
&\!\!=\!\!&
\frac{\sqrt{2}}{(3+k_{+})}\,\Big(
Q^2 V^6+Q^5 V^1
\Big)(z)\ , \ \  \ 
G^{-}_{\,\,2}(z)
=
\frac{\sqrt{2}}{(3+k_{+})}\,\Big(
Q^1 V^6-Q^6 V^1
\Big)(z)\ , 
\label{Gexp}
\eea
while the spin-$1$ currents associated to $\mathfrak{su}(2)_{k_+}$ 
are
\bea
A^{+1}(z)&=&-\frac{1}{2}\,\Big( V^3+V^7\Big)(z)\ ,
\nonu\\
A^{+2}(z)&=&-\frac{i}{2}\,\Big( V^3-V^7\Big)(z)\ ,
\nonu\\
A^{+3}(z)&=&\frac{1}{4}\,\Big((-i+\sqrt{3})\,V^4+(i+\sqrt{3})\,V^8 \Big)(z)\ .
\label{A+iexp}
\eea
The spin-$1$ currents corresponding to $\mathfrak{su}(2)_{k_-}$ on the other hand, are 
\bea
A^1(z)&=&-\frac{1}{2(3+k_{+})}\,\Big( Q^1Q^6+ Q^2 Q^5\Big)(z)\ ,
\nonu\\
A^2(z)&=&-\frac{i}{2(3+k_{+})}\,\Big( Q^1Q^6- Q^2 Q^5\Big)(z)\ ,
\nonu\\
A^3(z)&=&\frac{1}{2(3+k_{+})}\,\Big( Q^1Q^5- Q^2 Q^6\Big)(z)\ .
\label{Aexp}
\eea
The lowest non-trivial higher spin current (at $h=1$) is 
\be
\Phi^{(1)}_0(z)
=-\frac{k_{+}}{(3+k_{+})^2}\,\Big( Q^1 Q^5 +Q^2 Q^6\Big)(z)
+\frac{1}{2(3+k_{+})}\,\Big((3i+\sqrt{3}) V^4 +(-3i+\sqrt{3}) V^8\Big)(z)\ .
\label{higherspin1exp}
\ee
The spin $\frac{3}{2}$ generators of this multiplet are explicitly described as 
\bea
\Phi^{(1),+1}_\frac{1}{2}(z)
&=&\frac{\sqrt{2}}{(3+k_{+})}\,\Big(
-Q^1 V^5+Q^6 V^2
\Big)(z)\ ,
\nonu\\
\Phi^{(1),+2}_\frac{1}{2}(z)
&=&\frac{\sqrt{2}}{(3+k_{+})}\,\Big(
Q^2 V^5+Q^5 V^2
\Big)(z)\ ,
\nonu\\
\Phi^{(1),-}_{\frac{1}{2}\,\,\,\,\,\,\,1}(z)
&=&\frac{\sqrt{2}}{(3+k_{+})}\,\Big(
Q^2 V^6+Q^5 V^1
\Big)(z)\ ,
\nonu\\
\Phi^{(1),-}_{\frac{1}{2}\,\,\,\,\,\,\,2}(z)
&=&\frac{\sqrt{2}}{(3+k_{+})}\,\Big(
Q^1 V^6+Q^6 V^1
\Big)(z)\ ,
\label{higherspin3halfexp}
\eea
while at spin $2$ we have 
\bea
\Phi^{(1)}_1(z)
&=&
-\frac{1}{2(3+k_{+})^2}\,\Big(
(-i+\sqrt{3})(Q^1 Q^5 V^5+Q^2 Q^6 V^4) \nonu \\
& & \qquad \qquad \qquad +
(i+\sqrt{3})(Q^1 Q^5 V^8+Q^2 Q^6 V^8)
\Big)(z)
\nonu\\
&& + \, 
\frac{1}{2(3+k_{+})}\,\Big(
(-i+\sqrt{3})\partial V^4+
(i+\sqrt{3})\partial V^8
-4\,V^1 V^5 +4 V^2 V^6
\Big)(z)
\ ,
\nonu\\
\Phi^{(1),++}_1(z)
&=&
\frac{2}{(3+k_{+})^2}\,\Big(
Q^1 Q^5 V^3+ Q^2 Q^6 V^3
\Big)(z)
+\frac{2}{3+k_{+}}\,\Big(
2\,V^2 V^5-\partial V^3
\Big)(z)\ ,
\nonu\\
\Phi^{(1),--}_1(z)
&=&
\frac{2}{(3+k_{+})^2}\,\Big(
Q^1 Q^5 V^7+ Q^2 Q^6 V^7
\Big)(z)
+\frac{2}{3+k_{+}}\,\Big(
2\,V^1 V^6-\partial V^7
\Big)(z)\ .
\label{higherspin2exp}
\eea
These are in fact the generating fields of the coset algebra at $N=1$, i.e.\ all remaining fields can be obtained from them by normal ordered products. In particular, the remaining spin $2$ fields in this multiplet can be written as 
\bea
\Phi^{(1),1}_1(z)
&=&-\frac{1}{2(3+k_{+})^2}\,\Big( 
(3\,i+\sqrt{3})\,(Q^1Q^6V^4+Q^2 Q^5V^4) \nonu \\ 
&& \qquad + 
(-3\,i+\sqrt{3})\,(Q^1Q^6V^8+Q^2 Q^5 V^8)
\nonu\\
&& \qquad + 
2k_{+}\,(\partial Q^1 Q^6+\partial Q^2 Q^5 -Q^1 \partial Q^6-Q^2 \partial Q^5 )
\Big)(z)\ ,
\nonu\\
\Phi^{(1),2}_1(z)
&=&-\frac{i}{2(3+k_{+})^2}\,\Big( 
(3\,i+\sqrt{3})\,(Q^1Q^6V^4-Q^2 Q^5V^4) \nonu \\
& & \qquad +(-3\,i+\sqrt{3})\,(Q^1Q^6V^8-Q^2 Q^5 V^8) 
\nonu\\
&& \qquad + 
2k_{+}\,(\partial Q^1 Q^6-\partial Q^2 Q^5 -Q^1 \partial Q^6+Q^2 \partial Q^5 )
\Big)(z)\ ,
\nonu\\
\Phi^{(1),3}_1(z)
&=&\frac{1}{2(3+k_{+})^2}\,\Big( 
(3\,i+\sqrt{3})\,(Q^1Q^5V^4-Q^2 Q^6V^4) \nonu \\ 
& & \qquad 
+(-3\,i+\sqrt{3})\,(Q^1Q^5V^8-Q^2 Q^6 V^8)
\nonu\\
&& \qquad + 
2k_{+}\,(\partial Q^1 Q^5-\partial Q^2 Q^6 -Q^1 \partial Q^5+Q^2 \partial Q^6 )
\Big)(z)\, , 
\eea
while at spin $\frac{5}{2}$ we have 
\bea
\Phi^{(1),+1}_\frac{3}{2}(z)
&=&-\frac{\sqrt{2}}{(3+k_{+})^2}\,\Bigg( 
\frac{8}{3+k_{+}}\,(Q^1 Q^2 Q^6 V^5+Q^1 Q^5 Q^6 V^2)
\nonu\\
&& \quad +\,
4\,(Q^1 V^3 V^6+Q^6 V^1 V^3+i\,Q^6 V^2 V^4-i\,Q^6 V^2 V^8)
+2(i+\sqrt{3})\,Q^1 V^4 V^5
\nonu\\
&& \quad + \,
2(-i+\sqrt{3})\,Q^1 V^5 V^8
+\frac{8(2+k)}{3}\,(\partial Q^1 V^5-\partial Q^6 V^2)
\nonu\\
&& \quad +\, \frac{2(1-3i\sqrt{3}+2k_{+})}{3}\,Q^1 \partial V^5 + 
\frac{4(2+k_{+})}{3}\,Q^6 \partial V^2\Bigg)(z)\ ,
\nonu\\
\Phi^{(1),+2}_\frac{3}{2}(z)
&=&-\frac{\sqrt{2}}{(3+k_{+})^2}\,\Bigg( 
\frac{8}{3+k_{+}}\,(Q^1 Q^2 Q^5 V^5-Q^2 Q^5 Q^6 V^2)
\nonu\\
&& \quad -\, 
4\,(Q^2 V^3 V^6-Q^5 V^1 V^3-i\,Q^5 V^2 V^4+i\,Q^5 V^2 V^8)
-2(i+\sqrt{3})\,Q^2 V^4 V^5
\nonu\\
&& \quad -\, 
2(-i+\sqrt{3})\,Q^2 V^5 V^8
-\frac{8(2+k)}{3}\,(\partial Q^2 V^5+\partial Q^5 V^2)
\nonu\\
&& \quad +\, \frac{2(1-3i\sqrt{3}+2k_{+})}{3}\,Q^2 \partial V^5 + 
\frac{4(2+k_{+})}{3}\,Q^5 \partial V^2\Bigg)(z)\ ,
\nonu \\
\Phi^{(1),-}_{\frac{3}{2}\,\,\,\,\,\,\,1}(z)
&=&-\frac{\sqrt{2}}{(3+k_{+})^2}\,\Bigg( 
\frac{8}{3+k_{+}}\,(Q^1 Q^2 Q^5 V^6+Q^2 Q^5 Q^6 V^1)
\nonu\\
&& \quad -\, 
4\,(Q^2 V^5 V^7-Q^5 V^2 V^7-i\,Q^2 V^6 V^8+i\,Q^2 V^4 V^6)
-2(i+\sqrt{3})\,Q^5 V^1 V^4
\nonu\\
&& \quad -\, 
2(-i+\sqrt{3})\,Q^5 V^1 V^8
-\frac{8(2+k)}{3}\,(\partial Q^2 V^6-\partial Q^5 V^1)
\nonu\\
&& \quad +\, \frac{2(7+3i\sqrt{3}+2k_{+})}{3}\,Q^2 \partial V^6
- \frac{4(2+k_{+})}{3}\,Q^5 \partial V^1\Bigg)(z)\ , 
\nonu\\
\Phi^{(1),-}_{\frac{3}{2}\,\,\,\,\,\,\,2}(z)
&=&-\frac{\sqrt{2}}{(3+k_{+})^2}\,\Bigg( 
\frac{8}{3+k_{+}}\,(Q^1 Q^5 Q^6 V^1-Q^1 Q^2 Q^6 V^6)
\nonu\\
&& \quad -\, 
4\,(Q^1 V^5 V^7-Q^6 V^2 V^7-i\,Q^1 V^4 V^6-i\,Q^1 V^6 V^8)
+2(i+\sqrt{3})\,Q^6 V^1 V^4
\nonu\\
&& \quad +\, 
2(-i+\sqrt{3})\,Q^6 V^1 V^8
-\frac{8(2+k)}{3}\,(\partial Q^1 V^6+\partial Q^6 V^1)
\nonu\\
&& \quad +\, \frac{2(7+3i\sqrt{3}+2k_{+})}{3}\,Q^1 \partial V^6 +
\frac{4(2+k_{+})}{3}\,Q^6 \partial V^1\Bigg)(z)\ , 
\eea
and finally at spin $3$ we find 
\bea
\Phi^{(1)}_2(z)
&=&-\frac{8}{(3+k_{+})^4}\,\Bigg((3\,i+\sqrt{3})Q^1 Q^2 Q^5 Q^6 V^4-(3\,i-\sqrt{3})Q^1 Q^2 Q^5 Q^6 V^8
\\
&& -\, 
k_{+}\,Q^1 Q^2 \partial  (Q^5 Q^6) +\cdots
\Bigg)(z)
+\cdots
+\frac{4k_{+}(2+k_{+})}{3(3+k_{+})^3}\,\Big(\partial^2 Q^1 Q^5 +\partial^2 Q^2 Q^6 \Big)(z) \nonu
\eea
\bea
&& +\, 
\frac{(3i+5\sqrt{3}+3i\, k_{+}+\sqrt{3}k_{+})}{3(3+k_{+})^2}\,\partial^2 V^4(z)
+\frac{(-3i+5\sqrt{3}-3i\, k_{+}+\sqrt{3}k_{+})}{3(3+k_{+})^2}\,\partial^2 V^8(z)
\ .
\nonu
\eea
Here we have not written down the full expression for
the higher spin-$3$ field for simplicity. By substituting these expressions into (\ref{variphispin1}) --  (\ref{nullspin3}), and using the results of Section~\ref{sec:trunc},  we have checked that the null fields are indeed identically zero,
\bea
\varphi_{1}^{(1),i} =0\ , \qquad
\varphi_{\frac{3}{2}}^{(1),+a} =0\ ,
\qquad
\varphi_{\frac{3}{2}\,\,\,\,\,\,a}^{(1),-} =0\ ,
\qquad
\varphi_{2}^{(1)} =0\ .
\label{nullappendix}
\eea

\end{document}